\documentclass[prd,aps,twocolumn,preprintnumbers, showpacs, nofootinbib,superscriptaddress,notitlepage]{revtex4-2}
\usepackage{mathrsfs}
\usepackage{amsfonts}
\usepackage{amsmath}
\usepackage{slashed}
\usepackage{array}
\usepackage{verbatim}
\usepackage{epsfig}
\usepackage{graphicx}
\usepackage{color}
\usepackage[dvipsnames]{xcolor}
\usepackage[colorlinks,linkcolor=red,anchorcolor=green,citecolor=blue,CJKbookmarks=True]{hyperref}

\usepackage{multirow}
\usepackage{makecell}
\usepackage{booktabs}

\usepackage{ulem}

\newcommand{\beq}{\begin{eqnarray}}
\newcommand{\eeq}{\end{eqnarray}}

\newcommand{\MSbar}{{\overline{\text{MS}}}}

\def\be{\begin{equation}}
\def\ee{\end{equation}}
\def\bea{\begin{eqnarray}}
\def\eea{\end{eqnarray}}
\def\bal#1\eal{\begin{align}#1\end{align}}

\begin{document}

\title{Impact of Dynamical Charm Quark and Mixed Action Effect on Light Hadron Masses and Decay Constants}

\collaboration{\bf{CLQCD Collaboration}}

\author{
\includegraphics[scale=0.30]{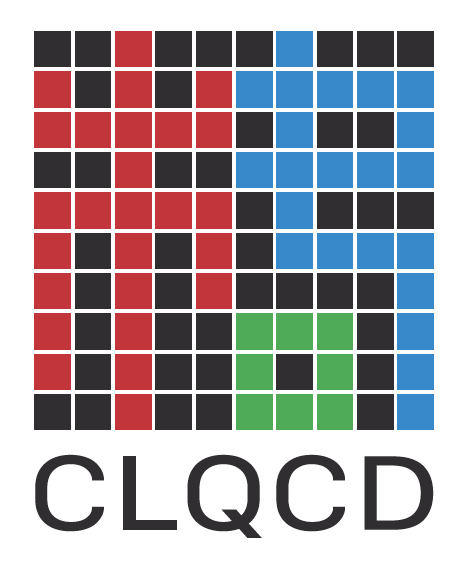}\\
Tong-Wei Lin}
\affiliation{School of Fundamental Physics and Mathematical Sciences, Hangzhou Institute for Advanced Study, UCAS, Hangzhou 310024, China}
\affiliation{Institute of Theoretical Physics, Chinese Academy of Sciences, Beijing 100190, China}
\affiliation{University of Chinese Academy of Sciences, Beijing 100049, China}

\author{
Zun-Xian Zhang}
\affiliation{School of Fundamental Physics and Mathematical Sciences, Hangzhou Institute for Advanced Study, UCAS, Hangzhou 310024, China}
\affiliation{Institute of Theoretical Physics, Chinese Academy of Sciences, Beijing 100190, China}
\affiliation{University of Chinese Academy of Sciences, Beijing 100049, China}

\author{Mengchu Cai}
\email[Corresponding author: ]{caimengchu@itp.ac.cn}
\affiliation{Institute of Theoretical Physics, Chinese Academy of Sciences, Beijing 100190, China}

\author{Hai-Yang Du}
\affiliation{Institute of Theoretical Physics, Chinese Academy of Sciences, Beijing 100190, China}
\affiliation{University of Chinese Academy of Sciences, School of Physical Sciences, Beijing 100049, China}

\author{Bolun Hu}
\affiliation{Computation-based Science and Technology Research Center, The Cyprus Institute, 20 Kavafi Str., Nicosia 2121, Cyprus}

\author{Xiangyu Jiang}
\affiliation{Department of Physics, Indiana University, Bloomington, Indiana 47405, USA}

\author{Xiao-Lan Meng}
\affiliation{Institute of Theoretical Physics, Chinese Academy of Sciences, Beijing 100190, China}
\affiliation{University of Chinese Academy of Sciences, School of Physical Sciences, Beijing 100049, China}

\author{Ji-Hao Wang}
\affiliation{Institute of Theoretical Physics, Chinese Academy of Sciences, Beijing 100190, China}
\affiliation{University of Chinese Academy of Sciences, School of Physical Sciences, Beijing 100049, China}

\author{Peng Sun}
\affiliation{Institute of Modern Physics, Chinese Academy of Sciences, Lanzhou, 730000, China}

\author{Yi-Bo Yang}
\email[Corresponding author: ]{ybyang@itp.ac.cn}
\affiliation{Institute of Theoretical Physics, Chinese Academy of Sciences, Beijing 100190, China}
\affiliation{University of Chinese Academy of Sciences, School of Physical Sciences, Beijing 100049, China}
\affiliation{School of Fundamental Physics and Mathematical Sciences, Hangzhou Institute for Advanced Study, UCAS, Hangzhou 310024, China}
\affiliation{International Centre for Theoretical Physics Asia-Pacific, Beijing/Hangzhou, China}

\author{Dian-Jun Zhao}
\affiliation{School of Science and Engineering, The Chinese University of Hong Kong, Shenzhen 518172, China}

\date{\today}

\begin{abstract}
We investigate the impact of including a dynamical charm quark on the properties of light hadrons. Our study compares the calculations performed on 2+1+1 flavor (HISQ fermion) ensembles at four lattice spacings to those on 2+1 flavor (clover fermion) ensembles at six lattice spacings, with both sets of ensembles employing the identical Symanzik gauge action. For the light, strange and charm flavor observables, we employ the same tadpole-improved clover fermion action. From numerical results for light and strange quark masses, pion and kaon decay constants, and $\Omega$ and $\Omega_{ccc}$ baryon masses, we find that the values obtained after continuum, chiral, and infinite-volume extrapolations are consistent within uncertainties. Even though the mixed action setup can introduce additional discretization effects, our calculation shows evidences that those effects can cancel with the discretization error in the unitary setup, resulting in better convergence in the continuum extrapolation. 
\end{abstract}

\maketitle

\section{Introduction}\label{sec:intro}

The Standard Model of particle physics includes six flavors of quarks. Three of these--charm, bottom, and top--are heavier than the intrinsic scale of the strong interaction, $\Lambda_{\rm QCD} \sim 300\ \mathrm{MeV}$. The other three--up, down, and strange--are lighter than $\Lambda_{\rm QCD}$ at a typical renormalization scale of 2 GeV in the $\overline{\mathrm{MS}}$ scheme.

The three light flavors exhibit an approximate $\mathrm{SU}(3)$ flavor symmetry, with symmetry-breaking effects on the order of $10\%$ for various observables. In contrast, the three heavy flavors are largely decoupled from light-hadron physics, a result supported by current lattice QCD averages in flavor physics~\cite{FlavourLatticeAveragingGroupFLAG:2024oxs}.

A direct assessment for the impact of heavy flavors on the light-hadron physics is highly non-trivial. Introducing an additional flavor significantly alters the lattice input value of $\beta$, which in turn modifies the dimensionless quark mass parameters that determine physical hadron masses. Furthermore, for widely used Wilson-like fermion formulations--such as clover and twisted-mass fermions--these bare masses require additional parameter tuning due to additive mass renormalization from explicit chiral symmetry breaking. In contrast, mass tuning for staggered fermions is much simpler. However, calculations of hadronic observables with staggered fermions are complicated by ambiguities arising from taste-breaking effects.

This paradox can be resolved by using a mixed action approach: gauge ensembles are generated with dynamical staggered fermions across various flavor numbers, while hadronic observables are calculated using fermion actions without taste degree of freedom such as clover. Although the leading-order low-energy constant for mixed action effects has been shown to scale as roughly \(\mathcal{O}(a^4)\) across a wide range of lattice spacings~\cite{Zhao:2022ooq,Zhang:2025ieg}, a systematic study of their impact on hadronic observables remains absent from the literature.

In this work, we compute light hadronic observables using clover valence fermions on dynamical 2+1+1 flavor highly improved staggered quark (HISQ) fermion and Symanzik gauge ensembles spanning four lattice spacings, three pion masses, and several spatial volumes. Compared to results obtained using the 2+1 flavor ensembles using the same clover fermion and Symanzik gauge actions, we find that the discretization errors from this mixed action setup are significantly reduced, with potential \(\mathcal{O}(a^4)\) contributions remaining negligible within our statistical uncertainties. Finally, we update the predictions for low-energy constants from a previous 2+1 flavor CLQCD study~\cite{CLQCD:2023sdb}, now with improved control over systematic uncertainties using the ensembles at three more lattice spacings.

\section{Simulation setup}\label{sec:setup}

In this work, we use the 1-step stout link smeared (with smear size $\rho=0.125$) clover valence fermion action~\cite{CLQCD:2023sdb,CLQCD:2024yyn},
\begin{align}
    \begin{aligned}
        &S_q(V, m) = \sum_{x} \bar{\psi}(x) \sum_{\eta=\pm 1}\sum_{\mu=1}^{4} \frac{\eta\gamma_\mu-1}{2} V_{\eta\mu}(x) \psi(x + \eta\hat{\mu} a)\\
    &+ \sum_x \psi(x) \left[ (4 + \tilde{m}) - c_{sw}\frac{a}{2} \sum_{\mu<\nu}\sigma^{\mu \nu} F_{\mu \nu}^V \right] \psi(x)\,,
    \end{aligned}
    \label{eq:quark_action}
\end{align}
where $c_{\rm sw}=1/v^3_{0}$ with $v_0$ being the tadpole-improved factor of the smeared gauge link $V$, and we use $\tilde{O}$ for the dimensionless value of any quantity $O$. $F_{\mu \nu}^{V}$ is defined as
\begin{align}
    \begin{aligned}
        F_{\mu \nu}^{V} &= \frac{i}{8 a^2}
    \left(
        \mathcal{P}_{\mu,\nu}^{V}
        - \mathcal{P}_{\nu,\mu}^{V}
        + \mathcal{P}_{\nu,-\mu}^{V}
        - \mathcal{P}_{-\mu,\nu}^{V}\right.\\
        &+ \left.\mathcal{P}_{-\mu,-\nu}^{V}
        - \mathcal{P}_{-\nu,-\mu}^{V}
        + \mathcal{P}_{-\nu,\mu}^{V}
        - \mathcal{P}_{\mu,-\nu}^{V}
    \right)\,,
    \end{aligned}
    \label{eq:SqVm}
\end{align}
with $\mathcal{P}^{V}_{\mu,\nu}(x) = V_{\mu}(x) V_{\nu}(x + a \hat{\mu}) V^{\dagger}_{\mu}(x + a \hat{\nu}) V^{\dagger}_{\nu}(x)$.

The calculation are performed on two sets of the gauge ensembles: one set is the 2+1 flavor ensembles using the tadpole-improved Symanzik gauge action and the same sea fermion action, and the other is the 2+1+1 flavor ones using the same gauge action but the HISQ fermion action. The first set has been used in the previous CLQCD works (e.g., those for the basic parameter calibrations~ \cite{CLQCD:2023sdb,CLQCD:2024yyn}), and the second set is generated recently and has been used for the determination of the mixed action effect in the valence-sea mixed pion mass~\cite{Zhang:2025ieg}. 

\begin{table}[!h]
\centering
\caption{Summary table on the lattice spacing, $m_{\pi}L$, $\tilde{L}^3\times \tilde{T}$, and $m_{\pi,\eta_s}$ of the CLQCD ensembles with clover fermion. The ensembles with superscript ${*}$ are newly generated after the previous CLQCD work~\cite{CLQCD:2023sdb}.}
\resizebox{1.0\columnwidth}{!}{
\begin{tabular}{l|ccc|cc}
\hline
Ensemble & $a$ (fm) &$m_{\pi}L$ &$\tilde{L}^3 \times \tilde{T}$ & $m_{\pi}\,(\mathrm{MeV})$ &$m_{\eta_s}\,(\mathrm{MeV})$ \\
\hline 
C24P34 &\multirow{6}{*}{0.10542(17)(62)} &4.37 &$24^3\times 64$ &340.6(1.7) &749.2(0.7)  \\
C24P29 & &3.75 &$24^3\times 72$ &292.4(1.0) &658.5(0.6)  \\
C32P29 & &5.02 &$32^3\times 64$ &293.4(0.8) &659.5(0.4)  \\
C32P23 & &3.90 &$32^3\times 64$ &228.1(1.2) &644.6(0.4)  \\
C48P23 & &5.75 &$48^3\times 96$ &224.3(1.2) &644.8(0.6)  \\
C48P14 & &3.50 &$48^3\times 96$ &136.5(1.7) &707.3(0.4)  \\
\hline
E28P35$^*$ &\multirow{3}{*}{0.09013(25)(53)} &4.42 &$28^3\times 64$ &345.4(1.1) &710.7(1.0)  \\
E32P29$^*$ & &4.17 &$32^3\times 64$ &285.4(1.8) &698.2(0.9)  \\
E32P22$^*$ & &3.14 &$32^3\times 96$ &215.1(2.2) &685.8(0.7)  \\
\hline
F32P30 &\multirow{5}{*}{0.07760(07)(46)} &3.79 &$32^3\times 96$ &301.0(1.2) &677.3(1.0)  \\
F48P30 & &5.73 &$48^3\times 96$ &303.5(0.7) &676.1(0.5)  \\
F32P21 & &2.65 &$32^3\times 64$ &210.5(2.2) &660.1(0.9)  \\
F48P21 & &3.93 &$48^3\times 96$ &207.9(1.1) &663.7(0.6)  \\
F64P14$^*$ & &3.41 &$64^3\times 128$ &135.6(1.2) &681.0(0.5)  \\
\hline
G36P29$^*$ &0.06895(17)(41) &3.72 &$36^3\times 108$ &295.7(1.1) &692.6(0.5)  \\
\hline
H48P32 &0.05235(11)(31) &4.03 &$48^3\times 144$ &316.1(1.0) &690.6(0.7)  \\
\hline
I64P31$^*$ &0.03761(08)(22) &3.81 &$64^3\times 128$ &312.2(1.6) &671.4(1.3)  \\
\hline
\end{tabular}
}
\label{tab:clover_ensembles}
\end{table}

The lattice spacing \(a\), the finite-volume parameter \(m_{\pi}L\), the lattice volume \(\tilde{L}^3 \times \tilde{T}\), pion mass, and $\eta_s$ mass for the gauge ensembles generated with the clover fermion action are listed in Table~\ref{tab:clover_ensembles}. $\eta_s$ is a fictitious pseudoscalar meson whose interpolating field is $\bar{s}\gamma^5 s$, with only connected insertions. For the ensembles marked with a superscript asterisk, we generate new quark propagators and correlation functions; for the remaining ensembles, we reuse the data from previous CLQCD work~\cite{CLQCD:2023sdb}. The first uncertainty of lattice spacings is the statistical uncertainty, and the second uncertainty is from the gradient flow parameter $w_0=0.17355(92)\,\mathrm{fm}$ from FLAG~\cite{FlavourLatticeAveragingGroupFLAG:2024oxs}. The lattice spacings are determined by the global fit method and the uncertainties are estimated from the model average analysis~\cite{Boccaletti:2024guq,Borsanyi:2020mff,BMW:2014pzb,Jay:2020jkz}. The fit ansatz and more details can be found in Ref.~\cite{Cai:2026xja}. 

\begin{table}[!h]
\centering
\caption{Summary table on the lattice spacings, $m_{\pi}L$, $\tilde{L}^3\times \tilde{T}$, and $m_{\pi,\eta_s,\eta_c}$ (in unit of MeV) of the CLQCD ensembles with HISQ fermion~\cite{Zhang:2025ieg}. The ensemble with superscript ${\dagger}$ is not used in the joint fit due to the limit statistics (c48P13) or different sea flavors (x/y/z24P31).}
\begin{tabular}{l|lcc|ccc}
\hline
Ensemble & $a$ (fm) &$m_{\pi}L$ & $\tilde{L}^3 \times \tilde{T}$ & $m_{\pi}$ &$m_{\eta_s}$ & $m_{\eta_c}$ \\
\hline 
c24P31s &\multirow{6}[2]{*}{0.1084(4)}&4.13&$24^3\times 48$&313(2)&745(3)&2.973(12)\\
c24P31 &                              &4.07&$24^3\times 48$&309(2)&687(3)& 2.972(12)\\
c32P31 &                              &5.44&$32^3\times 48$&310(1)&686(3)&2.972(12)\\
c24P22 &                              &2.94&$24^3\times 48$&223(2)&685(3)&2.970(12) \\
c32P22 &                              &3.87&$32^3\times 48$&220(1)&684(3) &2.970(12)\\
c48P13$^{\dagger}$ &                  &3.53&$48^3\times 48$&134(1)&683(3)& 2.970(12)\\
\hline
e32P31 &0.0867(4)                     &4.41&$32^3\times 64$&313(2)&694(3)&3.015(13)\\
\hline
g32P32 &\multirow{2}[2]{*}{0.0710(3)} &3.65&$32^3\times 64$&317(3)&692(3)&2.981(13)\\
g48P31 &                              &5.38&$48^3\times 64$&311(2)&691(3)&2.983(13)\\
\hline
h48P31 &0.0473(3)                     &3.60&$48^3\times 96$&313(3)&692(5)&2.947(19)\\
\hline
\hline
x24P31$^{\dagger}$&0.1114(6)          &4.35&$24^3\times 48$&321(2)&711(4)&-\\
y24P31$^{\dagger}$&0.1116(6)          &4.22&$24^3\times 48$&311(2)&688(4)&-\\
z24P31$^{\dagger}$&0.1128(7)          &4.35&$24^3\times 48$&317(2)&704(4)&-\\
\hline
\end{tabular}
\label{tab:hisq_ensembles}
\end{table}

Similar information for the newly generated gauge ensembles with the HISQ fermion action is provided in Table~\ref{tab:hisq_ensembles}. The lattice spacing determination procedure can be found in Ref.~\cite{Zhang:2025ieg}. The strange and charm quark masses are tuned to their physical values for most ensembles. The exception is the c24P31s ensemble, for which we deliberately vary the quark masses to investigate mistuning effects. We have also generated three additional 2+1 flavor ensembles to isolate the impacts of the charm sea quark from those of the mixed action setup, using the tree level Symanzik gauge action (x24P31), tadpole-improved Symanzik gauge action (y24P31, the same as the preivous 2+1 flavor CLQCD ensembles), and Iwasaki gauge action (z24P31). 

Due to the non-zero critical quark mass \( \tilde{m}_{\rm crit} \) required to vanish the pion mass in the clover fermion formulation, a direct definition of the renormalized quark mass \( m_q^R \) from the bare parameter \( \tilde{m}^b_q \) is ambiguous and should be replaced by the PCAC quark mass extracted from pseudoscalar quarkonium correlation functions which is expressed as~\cite{JLQCD:2007xff,CLQCD:2023sdb,CLQCD:2024yyn}:
\begin{align}
   m_q^{\rm PC}&=\left.\frac{m_{\rm PS}\sum_{\vec{x}}\langle A_4(\vec{x},t)P^\dag(\vec{0},0) \rangle}{2\sum_{\vec{x}}\langle P(\vec{x},t)P^\dag(\vec{0},0) \rangle}\right|_{t\rightarrow \infty}\label{eq:mass_v1}, 
\end{align}
with \( m_{\mathrm{PS}} \) denoting the corresponding pseudoscalar meson mass, $A_{\mu}=\bar{\psi}\gamma_{5}\gamma_{\mu}\psi$ and $P=\bar{\psi}\gamma_{5}\psi$. The renormalized quark mass is then given by \( m^R_q = (Z_A/Z_P) \, m^{\rm PC}_q \).

We compute the ratios of renormalization constants \( Z_A/Z_V \) and \( Z_A/Z_P \) for the clover action using amputated vertex functions in Landau gauge~\cite{Martinelli:1994ty,CLQCD:2023sdb}. These functions, \( \Lambda_{\mathcal{O}}(\mu^{\rm RI}) = S^{-1}(p_1) G_{\mathcal{O}}(p_1,p_2) S^{-1}(p_2)|_{p_1^2=p_2^2=q^2=(\mu^{\rm RI})^2} \) with $q\equiv p_1-p_2$, are derived from the quark propagator \( S(p_{1,2}) \) in the momentum space and the correlator \( G_\mathcal{O}(p_1,p_2) \) for an off-shell quark with momentum \( p_{1,2} \).

In addition to the $Z_V$ obtained from the vector current normalization of the hadron matrix element, the ratio \( Z_A / Z_V \), is obtained from the projection of vertex functions onto their tree level results,
 \begin{align}
\frac{Z_A}{Z_V} = \frac{\frac{1}{12q^2}\text{Tr}[q_{\mu}\Lambda^\mu_{V}(\mu^{\rm RI})q_{\nu}\gamma^\nu]}{\frac{1}{12q^2}\text{Tr}[q_{\mu}\Lambda^\mu_{A}(\mu^{\rm RI})q_{\nu}\gamma^\nu\gamma_5]},
\end{align}
and the $\mu^{\rm RI}$ dependence in the right hand side should cancel between denominator and numerator. 
Including this correction is essential for obtaining accurate values of \( f_{\pi,K} \) after the continuum extrapolation~\cite{CLQCD:2023sdb}. For the renormalized PCAC mass, \( Z_A/Z_P(\mu^{\rm RI}) \) is evaluated in the RI/SMOM scheme~\cite{Sturm:2009kb} at the scale \( \mu^{\rm RI} \),
\begin{align}
\frac{Z_A}{Z_P}(\mu) = \left. \frac{\frac{1}{12}\text{Tr}[\Lambda_{P}(\mu^{\rm RI})\gamma_5]}{\frac{1}{12q^2}\text{Tr}[q_{\mu}\Lambda^\mu_{A}(\mu^{\rm RI})q_{\nu}\gamma^\nu\gamma_5]} \right|_{q^2=(\mu^{\rm RI})^2},
\end{align}
which is free of the Goldstone pole~\cite{Liu:2013yxz}, and then perturbatively converted to the $\overline{\mathrm{MS}}$ scheme and evolved to the scale of 2 GeV~\cite{Bednyakov:2020ugu,Kniehl:2020sgo,Gracey:2022vjc,Baikov:2016tgj}. 

For the hadronic observable, we generate quark propagators using the Coulomb gauge-fixed wall source at multiple time slices for a range of quark masses around the unitary light and strange quark masses. The PCAC quark masses, as well as the masses and decay constants of the pseudoscalar mesons, are then extracted through a joint fit of correlation functions constructed from different interpolating fields, following the methodology detailed in previous CLQCD works ~\cite{CLQCD:2023sdb,CLQCD:2024yyn}.

\begin{table}[!h]
\centering
\caption{Summary of clover valence fermion mass parameters for the Clover and HISQ ensembles with \( m_{\pi} \sim 0.3 \) GeV at different lattice spacings. The table lists the slopes between the PCAC quark mass and the tadpole-improved bare quark mass, as well as the critical quark masses ($\tilde{m}_{\rm crit}$ in the original definition of fermion action, tadpole-improved value $\tilde{m}^{\rm tad}_{\rm crit}$ and the associated dimensionful mass $m^{\rm tad}_{\rm crit}$).}
\begin{tabular}{l|c|cccc}
\hline
 & $a$ (fm) & $k_m^{\rm tad}$ &$\tilde{m}_{\rm crit}$ & $\tilde{m}^{\rm tad}_{\rm crit}$ &$m^{\rm tad}_{\rm crit}$ (GeV) \\
\hline 
\multirow{6}[2]{*}{\makecell{2+1\\Clover\\sea}} &0.105 &0.841(1) &-0.28571(3) &-0.09619(4) &-0.180(1) \\
&0.090 &0.878(1) &-0.25753(5) &-0.07918(6) &-0.173(1) \\
&0.078 &0.914(2) &-0.23546(4) &-0.06618(4) &-0.168(1) \\
&0.069 &0.934(1) &-0.21982(2) &-0.05785(2) &-0.166(1) \\
&0.052 &0.973(1) &-0.18882(1) &-0.04294(1) &-0.162(1) \\
&0.038 &0.998(1) &-0.15940(1) &-0.03119(1) &-0.164(1) \\
\hline 
\makecell{2+1\\HISQ\\sea} & 0.112 &0.937(2)&-0.21379(9) & -0.06548(9)&-0.1158(2) \\
\hline
\multirow{4}[2]{*}{\makecell{2+1+1\\HISQ\\sea}} 
& 0.108&0.947(2)&-0.20403(7)&-0.06056(8)&-0.1102(1) \\
& 0.087&0.971(1)&-0.17179(4)&-0.04445(4)&-0.1012(1) \\
& 0.071&0.984(2)&-0.15141(4)&-0.03520(5)&-0.0979(1) \\
& 0.047&1.003(1)&-0.12081(1)&-0.02302(1)&-0.0961(1) \\
\hline
\end{tabular}
\label{tab:critical_mass}
\end{table}

In the light quark region, the PCAC quark mass $\tilde{m}^{\rm PC}$ depends on the bare quark mass linearly,
\begin{align}
\tilde{m}^{\rm PC}&=k_m (\tilde{m}^{\rm b}-\tilde{m}_{\rm crit})\nonumber\\
&=k^{\rm tad}_m (\tilde{m}^{\rm b,tad}-\tilde{m}_{\rm crit}^{\rm tad}),
\end{align}
where $\tilde{m}_q^{\rm b, tad}=\frac{4+\tilde{m}^{\rm b}_q}{v_0}-4$ is the tadpole-improved bare quark mass. The fit parameters $k^{\rm tad}$ and $\tilde{m}_{\rm crit}^{\rm tad}$ for the ensembles with $m_{\pi}\sim 0.3$ GeV are collected in Table~\ref{tab:critical_mass}. We also list the values of $\tilde{m}_{\rm crit}=(4+\tilde{m}_{\rm crit}^{\rm tad})v_0-4$ and $m_{\rm crit}^{\rm tad}=\tilde{m}_{\rm crit}^{\rm tad}/a$ for comparison.

As \( m^{\rm PC} \propto m_{\pi}^2 \), the critical mass \( m_{\rm crit} \equiv \tilde{m}_{\rm crit}/a \) measures explicit chiral symmetry breaking in the clover action. The loop correction of the Wilson term introduces an \(\mathcal{O}(\alpha_s)\) correction to the dimensionless bare mass \(\tilde{m}^{\rm b}\) (and also to \(\tilde{m}_{\rm crit}\)), which corresponds to an inherent \(\mathcal{O}(\alpha_s/a)\) power divergence in \(m_{\rm crit}^{\rm b}\). However, with the fully tadpole-improved clover term, \(\tilde{m}^{\rm tad}_{\rm crit}\) becomes nearly linear in \(a\), so that \(m^{\rm tad}_{\rm crit}\) exhibits only a mild dependence on the lattice spacing, suggesting that this divergence is strongly suppressed.

Table~\ref{tab:critical_mass} also shows that \( m^{\rm tad}_{\rm crit} \) is $\sim$40\% smaller on the 2+1+1 HISQ ensembles than on the 2+1 clover ensembles. Since a comparable reduction is seen on the 2+1 HISQ ensemble, we conclude the effect is due mainly to the difference between HISQ and clover quark actions in the gauge configurations, not the dynamical charm quark loop.

Finally, we also observe a significantly weaker lattice spacing dependence for the parameter \( k_m^{\rm tad} \) on the HISQ ensembles. This feature would relate to similar reduction in the discretization errors for hadronic observables, a point that will be addressed in the subsequent discussion.

Using propagators with a partially quenched valence quark mass \(m_l^{\mathrm{v}}\) on gauge ensembles with a light sea quark mass \(m_l^{\mathrm{s}}\), the partially quenched chiral perturbation theory (PQ\(\chi\)PT) suggests the following ansatz for the pion mass and decay constant~\cite{Sharpe:1997by,CLQCD:2023sdb}:
\bal
m^2_{\pi,{\rm vv}}&=\Lambda_{\chi}^2  2 y_{\rm v}\big\{1+\frac{2}{N_f}[(2 y_{\rm v}-y_{\rm s})\mathrm{ln} (2y_{\rm v})+(y_{\rm v}-y_{\rm s})]\nonumber\\
&\quad +2y_{\rm v} (2\alpha_8-\alpha_5)+2y_{\rm s} N_f (2\alpha_6-\alpha_4)\big\}\nonumber\\
&\quad \times\left[ 1+c^{\pi}_{L}e^{-m_{\pi}L}+c^{\pi}_{s}(m_{\eta_s}^2-m_{\eta_s,{\rm phys}}^2)  \right]\nonumber\\
&\quad \times(1+c^{\pi}_{a^2}a^2),\\
F_{\pi,{\rm vv}}&=F(1-\frac{N_f}{2}(y_{\rm v}+y_{\rm s})\mathrm{ln}(y_{\rm v}+y_{\rm s})+y_{\rm v}\alpha_5  + y_{\rm s}N_f\alpha_4)\nonumber\\
&\quad \times \left[1+d^{\pi}_{L}e^{-m_{\pi}L}+d^{\pi}_{s}(m_{\eta_s}^2-m_{\eta_s,{\rm phys}}^2) \right]\nonumber\\
&\quad \times(1+d^{\pi}_{a^2}a^2)
\eal
where \(N_f=2\) is the number of light flavors, \(\Lambda_{\chi}=4\pi F\) is the intrinsic scale of the chiral perturbation theory (\(\chi\)PT) with \(F\) being the pion decay constant in the chiral limit, and  
$y_{\mathrm{v/s}}=\frac{\Sigma \, m_l^{\mathrm{v/s}}}{F^2 \Lambda_{\chi}^2}$  
is the dimensionless expansion parameter of \(\chi\)PT. Here, \(\Sigma\equiv -\langle \bar{q}q\rangle_{m_l\rightarrow 0}\) denotes the chiral condensate in the \(N_f=2\) chiral limit, and \(\alpha_{4,5,6,8}\) are the next-to-leading-order (NLO) low-energy constants. Additional parameters \(c^{\pi}_{L,s,a^2}\) and \(d^{\pi}_{L,s,a^2}\) are introduced to parameterize corrections from finite volume, unphysical strange-quark masses, and finite lattice spacings. The $m_{\pi}$ and $f_{\pi}$ with unitary valence and sea quark masses $y=y_{\rm v}=y_{\rm s}$ have another widely used parameterization,
\begin{align}
    m_\pi^2&=\Lambda_{\chi}^2  2 y[1+y(\ln\frac{2y\Lambda^2_{\chi}}{m^2_{\pi, {\rm phys}}}-{\ell}_3)+\mathcal{O}(y^2)],\\
    F_\pi&=F[1-2y(\ln\frac{2y\Lambda^2_{\chi}}{m^2_{\pi, {\rm phys}}}-\ell_4)+\mathcal{O}(y^2)].
\end{align}
where $\ell_{3,4}$ is related to $\alpha_{4,5,6,8}$ by
\bal\label{eq:NLO_LEC}
\ell_3=&\ln\frac{\Lambda^2_{\chi}}{m_{\pi,{\rm phys}}^2}-2[(2\alpha_8-\alpha_5)+2(2\alpha_6-\alpha_4)],\nonumber\\
\ell_4=&\ln\frac{\Lambda^2_{\chi}}{m_{\pi,{\rm phys}}^2}+\frac{1}{2}(\alpha_5+2\alpha_4).
\eal

Similarly, we fit the partially quenched kaon masses and decay constants on all the ensembles using the following form proposed in recent works~\cite{ExtendedTwistedMass:2021gbo,CLQCD:2023sdb},
\bal
&m_K^2(m^{\rm v}_l,m^{\rm s}_l,m^{\rm v}_s,m^{\rm s}_s,a,1/L) \nonumber\\
&\quad =(b_s^{\rm v} m^{\rm v}_s+b_s^{\rm s} m^{\rm s}_s+b_l^{\rm v} m^{\rm v}_l+b_l^{\rm s} m^{\rm s}_l) \notag \\ 
&~~~~~~ \times\left[ 1+c^K_l m^{\rm v}_l+c^K_{a^2} a^2 + c_{L}^K \exp{(-m_{\pi} L)} \right],\label{eq:kaon_mass}\\
&f_K(m^{\rm v}_l,m^{\rm s}_l,m^{\rm v}_s,m^{\rm s}_s,a,1/L)\nonumber\\
&\quad=(f_0+df_s^{\rm v} m^{\rm v}_s+df_s^{\rm s} m^{\rm s}_s+df_l^{\rm v} m^{\rm v}_l+df_l^{\rm s} m^{\rm s}_l) \notag\\
&~~~~~~ \times\left[1+d^K_{a^2} a^2 +d_{L}^K \exp{(-m_{\pi} L)}   \right].\label{eq:kaon_dc}
\eal
We further introduce an additional $a\alpha_s$ term in the fit ansatz of $f_{\pi,K}$ and take the difference as a systematic uncertainty, to capture the residual $a\alpha_s$ correction, where the value of $\alpha_s$ is approximated using the relation $\alpha^{u_0}_s\equiv -\frac{4}{3.06839} \log u_0$ for the Symanzik gauge action~\cite{Alford:1995hw,Zhang:2025ieg}. The joint fit of the pion mass and decay constant shall also introduce modification on the light quark mass due to the $a\alpha_s$ correction of $f_{\pi}$. The physical light and strange quark masses are then determined by imposing the isoQCD conditions \( m_{\pi} = 135.0(2) \) MeV and \( m_K = 494.6(1) \) MeV~\cite{DiCarlo:2019thl}.

At the same time, the masses of the $\phi$, $\Omega$ hadrons—which contain only valence strange quarks—are fitted using following empirical ansatz,
\begin{align}
 &m_{H=\phi/\Omega_{(ccc)}}(m^{\rm s}_l,m^{\rm v}_s,m^{\rm s}_s,a,1/L)=[m_{H}^{\rm phys}\nonumber\\
&\quad
+\sum_{m_i=m_{\pi}^{\rm sea},m_{\eta_s}^{\rm val/sea}}c_i^{H} (m^{2}_{i}-m^{2}_{i,{\rm phys}})] \times\left( 1+c^H_{a^2} a^2  \right).\label{eq:joint_fit_mh}
\end{align}
The corresponding values of $m_\pi^{\mathrm{sea}}$ on different ensembles required in the fit for $m_H$ are from Table~\ref{tab:hisq_ensembles}, which are calculated with the unitary quark mass with the HISQ action.

\section{Results}\label{sec:setup}


In this work, computations on the HISQ ensembles are performed using 50 configurations per ensemble, with 8 Coulomb wall source propagators generated on each configuration. For the illustration of the lattice spacing dependence of kinds of hadrons, we tune the valence strange quark mass to its physical value using \(m_{\eta_s}^{\mathrm{phys}} = 689.89(49)\,\text{MeV}\) from Ref.~\cite{Borsanyi:2020mff}. In Ref.~\cite{CLQCD:2023sdb}, the unitary clover fermion setup was used, and the physical strange quark mass was determined from the QED-subtracted kaon mass, yielding \(m_{\eta_s} = 687.4(2.2)\) MeV in the continuum limit. This value is consistent with that of Ref.~\cite{Borsanyi:2020mff} within uncertainties, indicating that the determination of the strange quark mass is independent of the specific fermion action employed. An alternative condition for determining the physical strange quark mass is
\begin{align}
    \sqrt{m_{K^+,{\rm QCD}}^2 + m_{K^0,{\rm QCD}}^2 - m_{\pi^+,{\rm QCD}}^2} = 686.1(1)~\mathrm{MeV},
\end{align}
based on the QED subtracted light meson masses~\cite{Giusti:2017dmp}. Such a condition should also yield a consistent determination.

\begin{figure}[hbt!]
\centering
\includegraphics[width=1.0\linewidth]{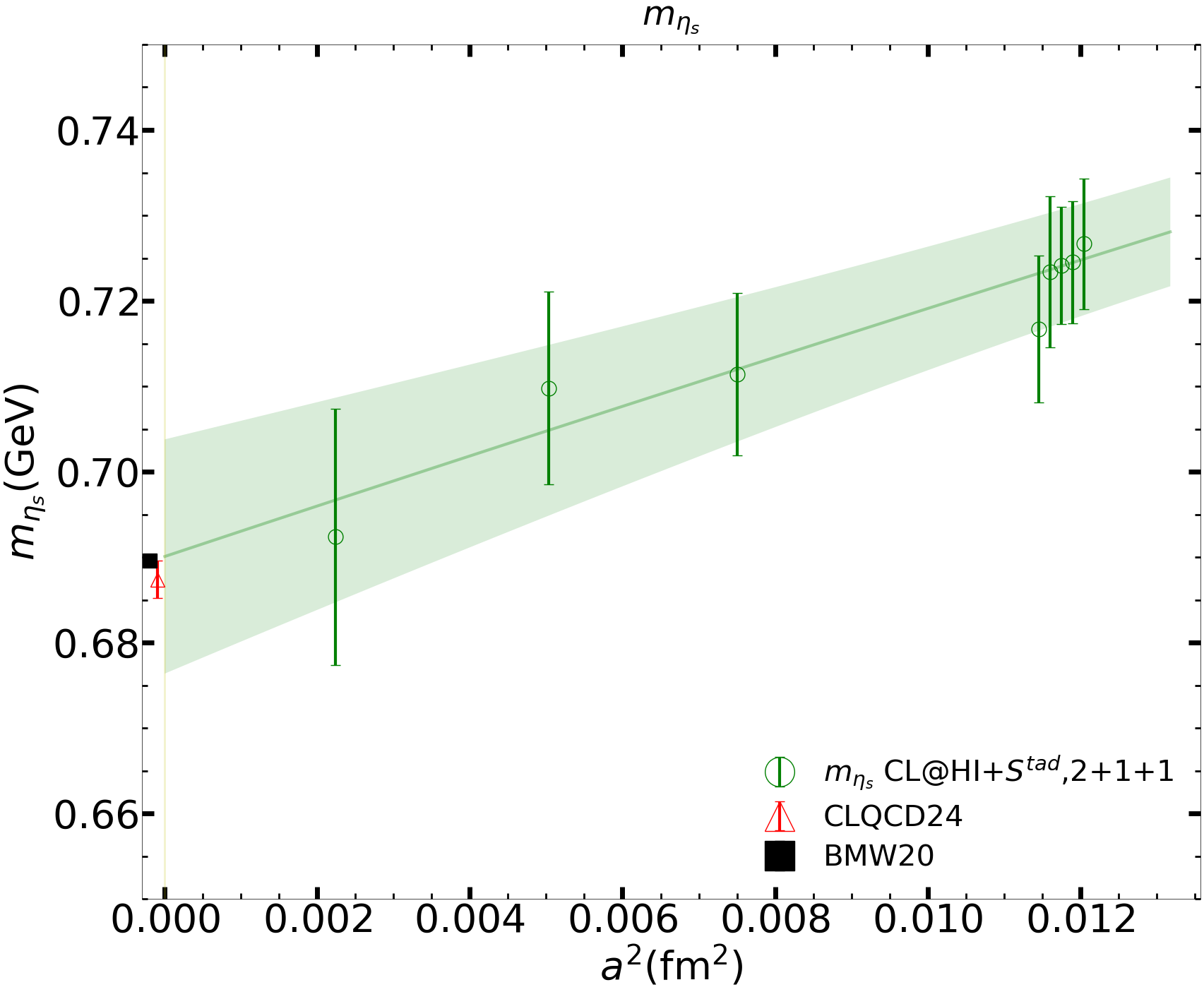}
\caption{Lattice spacing dependence of $m_{\eta_s}$ using the valence strange quark mass tuned according to the physical $m_{\phi}$, with the light and strange sea quark masses extrapolated to their physical values.}\label{fig:eta_s}
\end{figure}

As further justification, we also tune the valence strange quark mass using the physical \(\phi\) meson mass \(m_{\phi}^{\mathrm{phys}} = 1019.46(2)\) MeV~\cite{ParticleDataGroup:2024cfk} and then perform a joint fit for $m_{\eta_s}$ using the ansatz defined in Eq.~(\ref{eq:joint_fit_mh}). As illustrated in Fig.~\ref{fig:eta_s}, after applying quark mass corrections to the physical point for both the light and strange quarks based on the joint fit, the continuum-extrapolated value of \(m_{\eta_s}\) is \(691(14)\) MeV. This result is consistent with the values found in the literature~\cite{Borsanyi:2020mff,CLQCD:2023sdb}, albeit with a significantly larger uncertainty due to the relatively larger uncertainty in \(m_{\phi}\).

\begin{figure}[hbt!]
\centering
\includegraphics[width=1.0\linewidth]{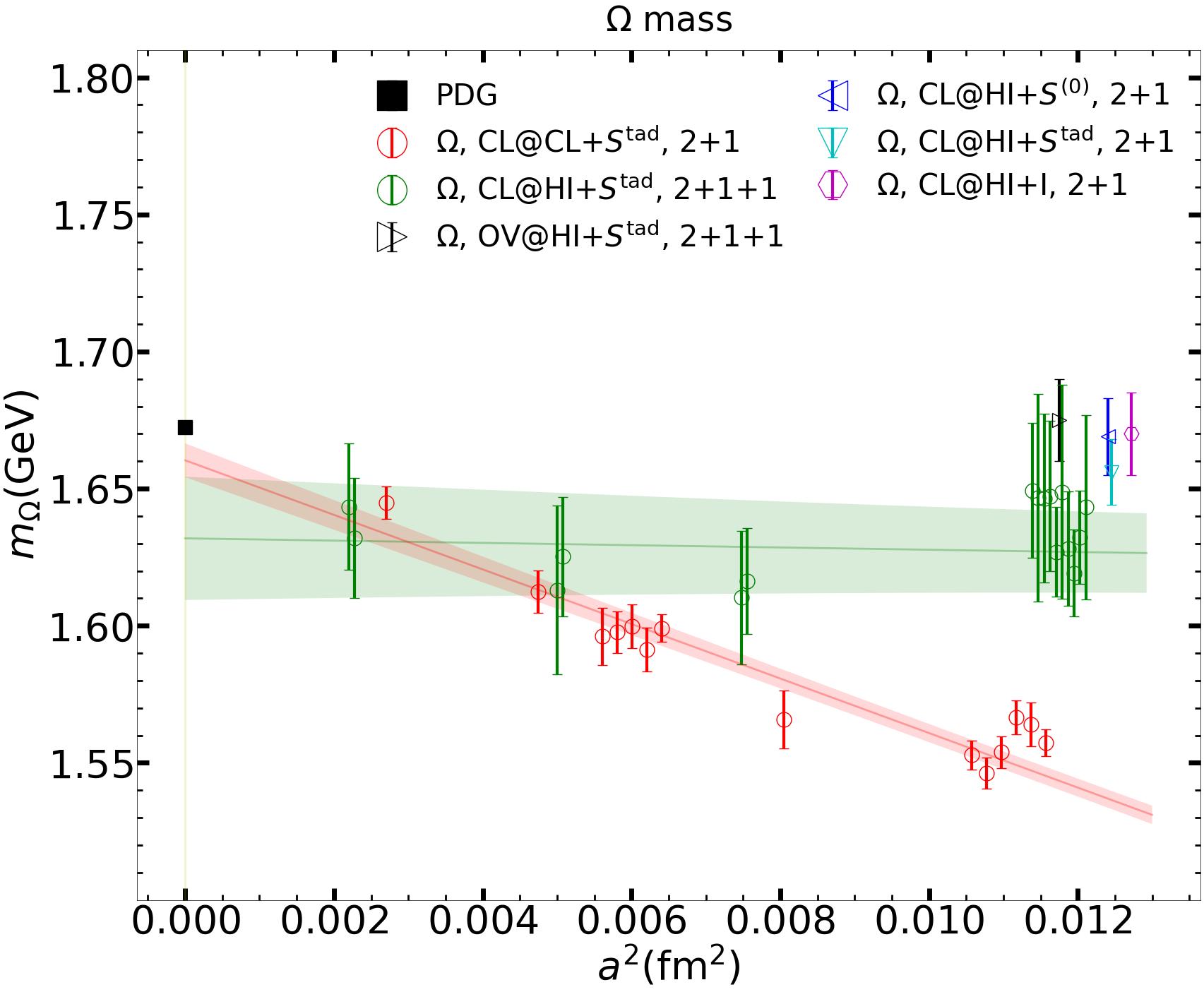}
\includegraphics[width=1.0\linewidth]{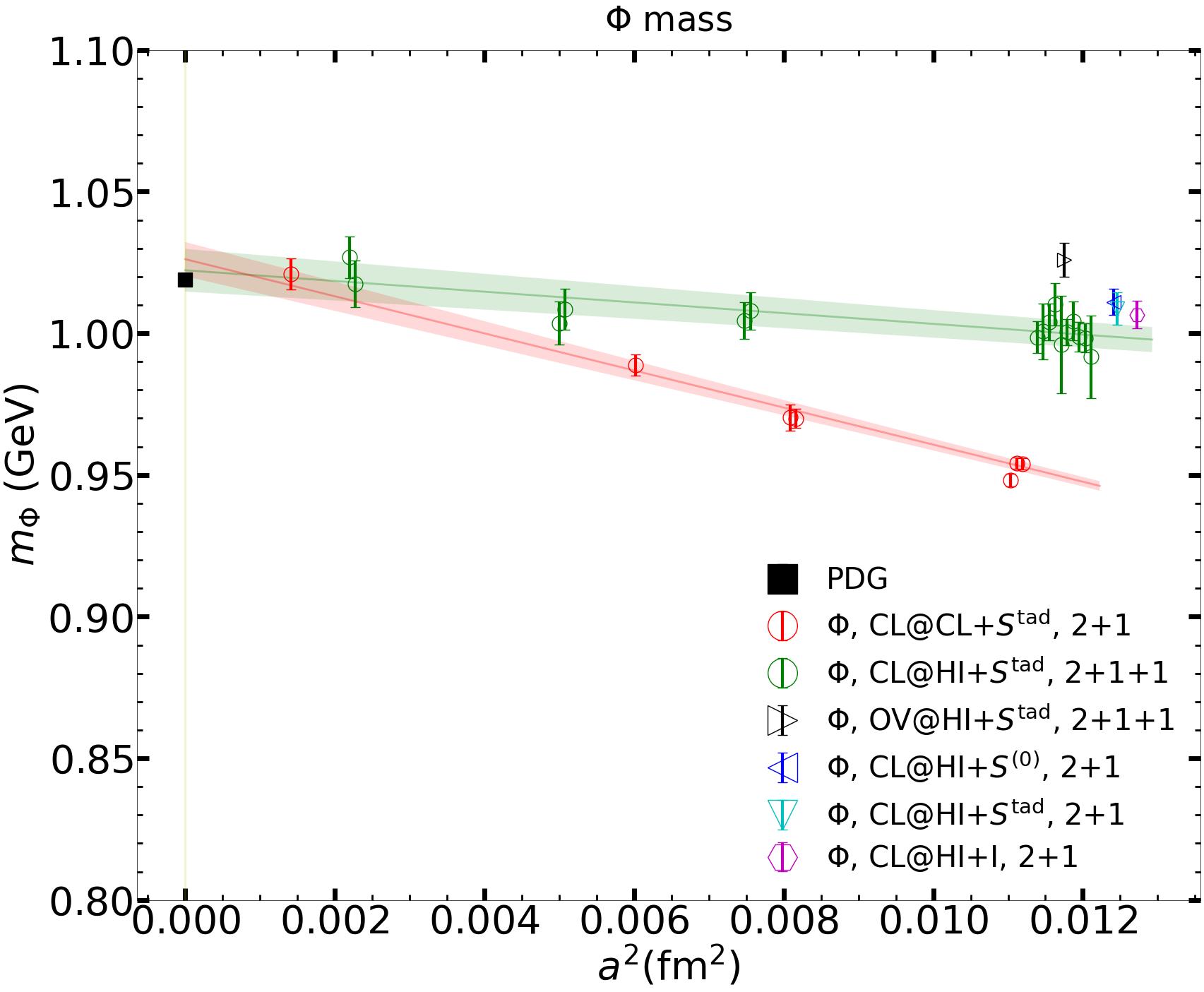}
\caption{Lattice spacing dependence of $m_{\Omega}$ (upper panel)
and $m_{\phi}$ (lower panel) with the quark masses extrapolated to their physical values.}\label{fig:strange}
\end{figure}

Then we compute the masses \(m_\phi\) and \(m_\Omega\) on \(N_f=2+1+1\) ensembles at four lattice spacings and various light-quark masses. The results, shown as green circles (labeled CL@HI) in Fig.~\ref{fig:strange}, are compared with analogous results from \(N_f=2+1\) ensembles using the unitary clover action~\cite{Hu:2024mas} (red points, labeled CL@CL). Despite the relatively larger uncertainties in the CL@HI results, the discretization errors for $m_{\phi}$ and \(m_\Omega\) are significantly smaller than those in the CL@CL case. It is observed that the PDG value of \( m_\Omega \) lies above most \( N_f = 2+1+1 \) data points, as well as our extrapolated results, despite the large statistical uncertainties. A possible explanation is that the quark mass dependence extracted from the global fit is primarily constrained by data from ensembles with the largest lattice spacings. This leads to \( m_\Omega \) values that, after applying quark mass corrections to the physical point, are generally lower. The quark mass dependence on finer lattice spacings may differ, a possibility that warrants further investigation in future work.

To isolate the effect of a dynamical charm sea, we performed similar calculations on three \(N_f=2+1\) HISQ ensembles (x/y/z24P31) at a comparable pion mass and lattice spacing (\(a \sim 0.11\ \text{fm}\)) but with different gauge actions. All three results agree well with each other and with those from the \(N_f=2+1+1\) ensembles. This indicates that the additional discretization error introduced by the valence-sea fermion mismatch can partially cancel the error inherent to the valence fermion action itself, leading to a reduced total discretization error—an effect that appears insensitive to the choice of gauge action or the inclusion of a charm sea.

\begin{figure}[hbt!]
\centering
\includegraphics[width=1.0\linewidth]{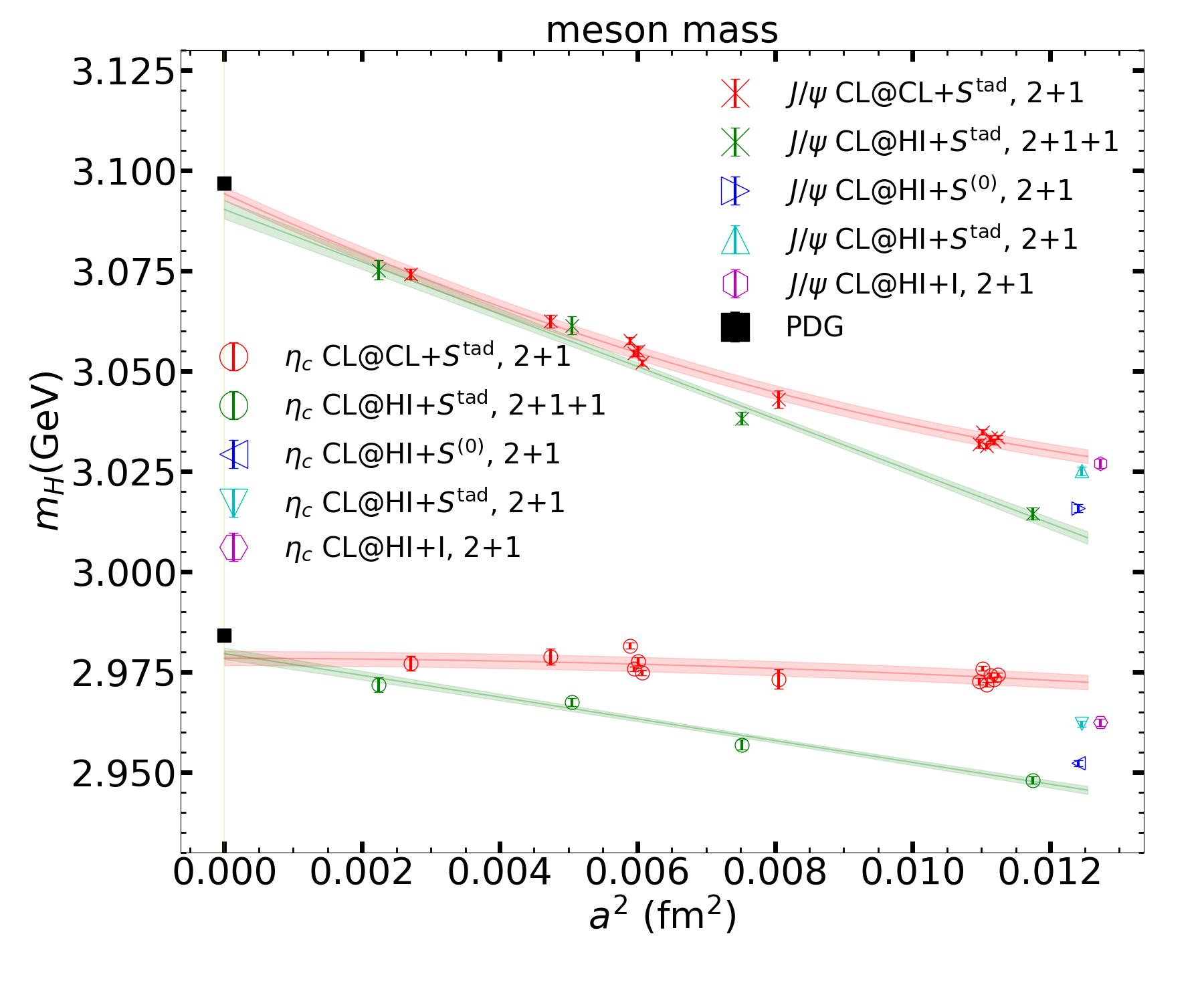}
\includegraphics[width=1.0\linewidth]{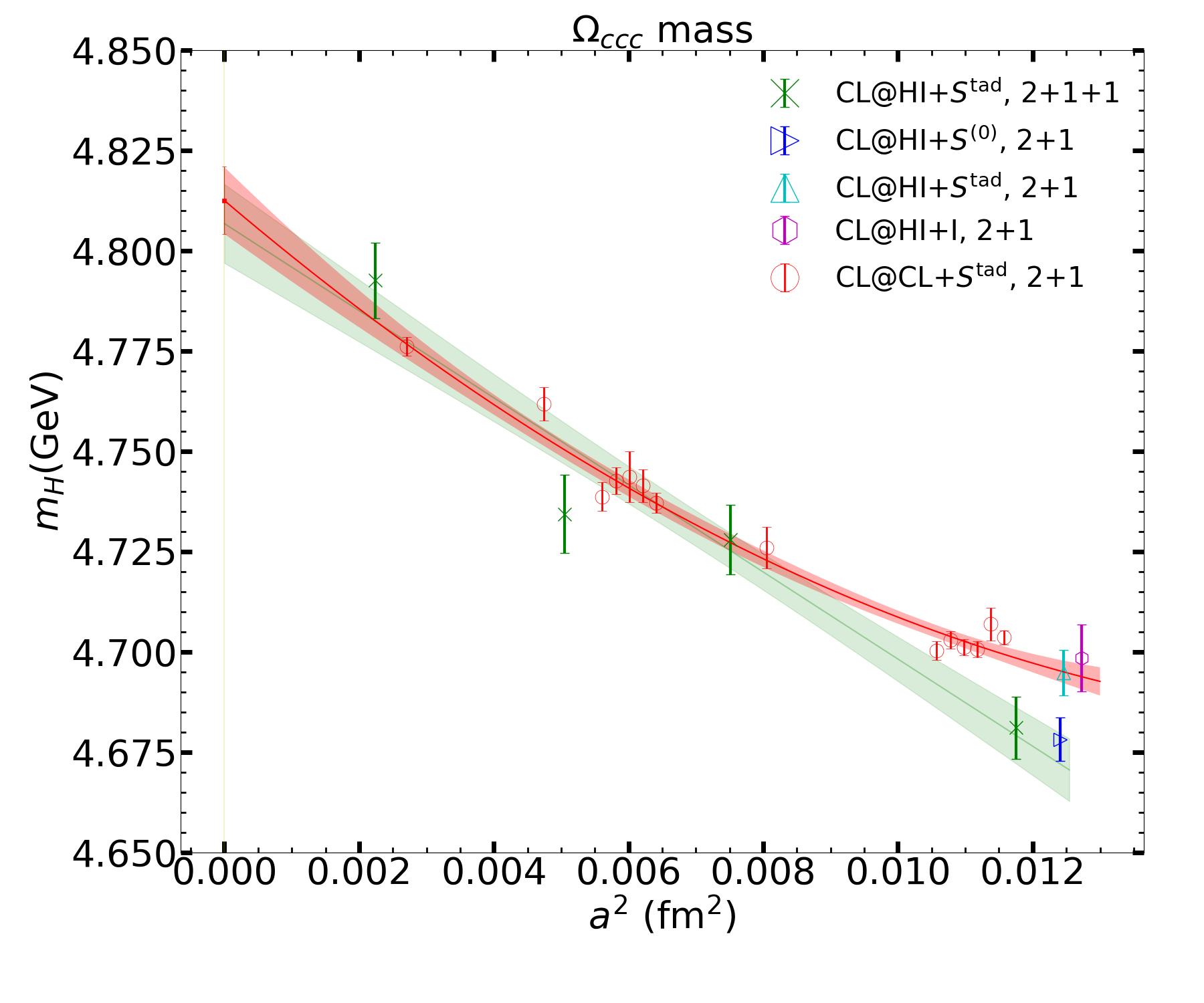}
\caption{Lattice spacing dependence of $m_{\eta_c}$ , $m_{J/\psi}$ (upper panel) and $m_{\Omega_{ccc}}$ (lower panel), using the valence charm quark mass tuned to the QED-correction-subtracted physical $m_{D_s}=1966.7(1.5)$ MeV.}\label{fig:charmoninum}
\end{figure}

Separately, using a valence charm quark mass tuned to the QED-correction-subtracted \(m_{D_s}=1966.7(1.5)\) MeV, we compute the S-wave charmonium masses \(m_{\eta_c}\) and \(m_{J/\psi}\) on the \(N_f=2+1+1\) ensembles c24P31 and g32P32 at two lattice spacings with \(m_\pi \sim 310\ \text{MeV}\). These results are shown as green crosses and circles (CL@HI) in the upper panel of Fig.~\ref{fig:charmoninum}. Compared to the corresponding unitary clover (\(N_f=2+1\)) results~\cite{CLQCD:2024yyn} (red points, CL@CL), the clover-on-HISQ data exhibit a larger discretization error for \(m_{\eta_c}\) but a smaller \(a^4\) correction for \(m_{J/\psi}\). In the continuum limit, the unitary clover results are consistent with clover-on-HISQ results, but they are slightly lower than the experiment values from PDG. This discrepancy may arise from the omission of disconnected diagrams in the calculations for $m_{J/\psi}$ and $m_{\eta_c}$.

To investigate the charm-sea effect here, we repeated the calculation on three \(N_f=2+1\) HISQ ensembles (x/y/z24P31) at similar parameters. On ensemble y24P31, which uses the same tadpole-improved Symanzik gauge action as our main sets, \(m_{J/\psi}\) agrees with the unitary-clover results, while \(m_{\eta_c}\) is lower, producing a hyperfine splitting \(\Delta^c_{\mathrm{HFS}} \equiv m_{J/\psi} - m_{\eta_c}\) similar to the \(N_f=2+1+1\) case. Results on z24P31 (Iwasaki gauge action) are comparable. In contrast, x24P31 (tree-level Symanzik action) yields significantly different individual masses, though \(\Delta^c_{\mathrm{HFS}}\) remains statistically consistent across all HISQ ensembles, regardless of gauge action or charm sea.

The lattice spacing dependence for the mass of $\Omega_{ccc}$ is shown in the lower panel of Fig.~\ref{fig:charmoninum}. In the continuum limit, the unitary clover (\(N_f=2+1\)) result is consistent with that from clover-on-HISQ (with tadpole-improved Symanzik gauge action) calculation, while the unitary clover result exhibits a larger $a^4$ discretization error. We also calculate $m_{\Omega_{ccc}}$ on three \(N_f=2+1\) HISQ ensembles (x/y/z24P31). The behaviors of the results are similar to those observed for $J/\psi$. Results on y24P31 and z24P31 are consistent with the extrapolated band of unitary clover calculation, whereas the result on x24P31 agrees with the extrapolated band of clover-on-HISQ case.

We therefore conclude that the discretization of light-fermion loops in the gauge ensemble has a significant impact on charmonium mass discretization errors. Meanwhile, given the much higher statistical precision achievable in the charm sector, the effects of the dynamical charm sea and the choice of gauge action are also non-negligible.

\begin{table*}[ht!]                   
    \caption{Summary of our determination on quark masses at $\MSbar{\rm (2 GeV)}$ and the other quantities, with comparison with FLAG~\cite{FlavourLatticeAveragingGroupFLAG:2021npn,FlavourLatticeAveragingGroupFLAG:2024oxs,RBC:2012cbl,CLQCD:2023sdb,Bruno:2019vup,RBC:2014ntl,Laiho:2011np,BMW:2010ucx,BMW:2010skj,Bazavov:2010yq,McNeile:2010ji,RBC:2010qam,Davies:2009ih,MILC:2009ltw,MILC:2009mpl,Mason:2005bj,MILC:2004qnl,HPQCD:2004hdp,Bruno:2019xed,ExtendedTwistedMass:2021gbo,EuropeanTwistedMass:2014osg,FermilabLattice:2018est,Lytle:2018evc,Chakraborty:2014aca,Bazavov:2017lyh,FermilabLattice:2014tsy,ExtendedTwistedMass:2021qui,Miller:2020xhy,Carrasco:2014poa,Dimopoulos:2013qfa,Dowdall:2013rya,MILC:2013ffd,MILC:2011nja,Farchioni:2010tb,QCDSF-UKQCD:2016rau,Durr:2016ulb,Scholz:2016kcr,MILC:2010hzw,BMW:2010xmi,Aubin:2008ie,Follana:2007uv,Fahy:2015xka,Liang:2021pql,Alexandrou:2017bzk,Cichy:2013gja,Aoki:2017paw,Boyle:2015exm,BMW:2013fzj,Borsanyi:2012zv,Engel:2014eea,Brandt:2013dua,Burger:2012ti,ETM:2009ztk,Frezzotti:2008dr,ETM:2010cqp,Beane:2011zm,Horsley:2013ayv,Colangelo:2003hf,Gulpers:2015bba,JLQCD:2009ofg} and/or PDG~\cite{ParticleDataGroup:2024cfk}. A second source of uncertainty arises in certain cases from the deviation in the central value when the $a\alpha_s$ terms are included in the joint fits of $f_{\pi}$ and $f_K$. This deviation can lead to a slight shift in the determined value of $m_l$, as the fits for $m_{\pi}$ and $f_K$ both depend on the same low-energy constants, $\Sigma$ and $F$. }  
    \begin{tabular}{l| ccccc | ccc }       
    \hline
    \hline 
    &\multicolumn{5}{c|}{CL@CL, 2+1}& & CL@HI, 2+1+1 & \\
\multirow{2}{*}{} & \multicolumn{2}{c}{MOM} & \multicolumn{2}{c}{SMOM} & \multirow{2}{*}{FLAG/PDG} &  \multirow{2}{*}{MOM} & \multirow{2}{*}{SMOM} & \multirow{2}{*}{FLAG/PDG} \\ 

 & Ref.~\cite{CLQCD:2023sdb} & This work & Ref.~\cite{CLQCD:2023sdb} & This work & & & & \\ 
\hline 
  $m_l$ (MeV)&3.60(11)&3.51(08)(00)& 3.45(05)&3.37(04)(00)&3.39(04)(00)&3.46(16)(00)&3.50(09)(00)&3.43(05) \\
  $m_s$ (MeV)&98.8(2.9)&97.5(1.7)&94.1(1.2)&93.4(1.0)&92.4(1.0)&96.8(4.7)&97.7(2.3)&93.46(58) \\
   $m_s/m_l$&27.47(30)&27.77(26)(01)&27.28(22)&27.68(20)(00)& 27.42(12)&28.00(79)(00)&27.89(60)(00)&27.23(08)  \\
 \hline
 $\Sigma^{1/3}$(MeV)&268.6(3.6)&268.8(2.7)(0.5)&269.3(1.8)&272.1(2.0)(0.9)&272(5)&275.8(6.3)(0.1)&275.8(4.7)(0.1)&286(23)\\
 $F $ (MeV)&86.6(0.7)&85.6(0.8)(0.3)&85.1(0.6)&85.5(0.9)(0.4)&&88.0(2.3)(0.1)&88.8(2.5)(0.0)&\\
 $\ell_3$&2.43(54)&2.58(35)(00)&2.49(23)&2.51(15)(00)&3.07(64)&3.57(92)(00)&2.98(56)(00)&3.53(26)  \\
 $\ell_4$&4.32(08)&4.25(05)(00)&4.23(05)&4.23(04)(00)&4.02(45)&4.02(13)(00)&4.02(13)(00)&4.73(10)  \\
 \hline 
 $f_{\pi}$(MeV)&130.7(0.9)&129.3(1.0)(0.3)&128.6(0.8)&129.0(1.0)(0.6)&130.2(0.8)&131.8(2.7)(0.1)&132.8(3.0)(0.1)\\
 $f_{K^{\pm}}$(MeV)&155.6(0.8)&156.3(0.9)(1.2)&152.9(0.7)&156.4(0.9)(1.1)&155.7(0.7)&157.3(2.1)(0.1)&158.2(2.7)(0.1)&155.7(0.3)\\
$F_{\pi}/F$&1.0675(19)&1.0674(17)(03)&1.0683(15)&1.0668(17)(04)&1.0620(70)&1.0593(59)(01)&1.0573(65)(01)&1.0770(30) \\
 $f_{K}/f_{\pi}$&1.1907(76)&1.2092(72)(57)&1.1890(74)&1.2123(73)(34)&1.1916(34)&1.194(18)(00)&1.192(17)(01)&1.1934(19) \\
 \hline
 $c_s^\pi$(GeV$^2$)&&0.07(11)(00)&&0.01(05)(00)&&-0.12(29)(00)&0.02(16)(00)& \\
 $d_s^\pi$(GeV$^2$)&&0.20(03)(00)&&0.19(03)(00)&&0.16(10)(00)&0.17(12)(00)& \\
 $b_s^{\rm v}$(GeV) &&2.34(08)&&2.50(03)&&2.58(22)&2.42(13)&\\
 $b_s^{\rm s}$(GeV) &&0.06(08)(00)&&0.01(04)(00)&&-0.16(30)&-0.02(16)&\\
 $b_l^{\rm v}$(GeV) &&2.33(41)&&2.26(19)&&4.2(1.7)&1.6(1.0)&\\
 $b_l^{\rm s}$(GeV) &&0.43(12)&&0.55(06)&&0.34(46)&0.51(21)&\\
 $c_l^{K}$(GeV$^{-1}$) &&0.8(1.3)&&0.7(0.6)&&-5.0(4.6)&3.3(3.8)&\\ 
 $df_s^{\rm v}$ &&0.184(06)(03)&&0.197(05)(03)&&0.165(23)(01)&0.176(20)(00)&\\
 $df_s^{\rm s}$ &&0.067(27)(04)&&0.069(22)(03)&&0.049(64)(00)&0.065(73)(00)&\\
 $df_l^{\rm v}$ &&0.225(19)(02)&&0.225(17)(01)&&0.133(62)(00)&0.176(61)(00)&\\
 $df_l^{\rm s}$ &&0.433(29)(01)&&0.416(27)(01)&&0.307(76)(00)&0.260(84)(01)&\\ 
 \hline 
 $c_{a^2}^\pi$(fm$^{-2}$)&&-0.2(3.2)&&-4.3(0.6)&&1.8(2.3)&-2.7(1.0)& \\
 $d_{a^2}^\pi$(fm$^{-2}$)&&-5.19(47)&&-3.98(41)&&0.74(71)&1.30(77)& \\
 $c_{a^2}^{K}$(fm$^{-2}$) &&-0.4(3.1)&&-4.17(53)&&1.0(2.0)&-3.5(1.1)&\\ 
 $d_{a^2}^{K}$(fm$^{-2}$) &&-6.02(65)&&-5.31(33)&&0.02(68)&0.09(72)&\\ 
 \hline
 $c_L^\pi$&&0.60(32)(00)&&0.51(19)(00)&&0.68(47)(00)&0.77(36)(00)& \\
 $d_L^\pi$&&-0.60(12)(00)&&-0.61(12)(00)&&-0.39(21)(00)&-0.34(22)(00) \\
 $c_L^{K}$ &&0.27(20)&&0.177(94)&&-0.06(37)&0.23(21)&\\ 
 $d_L^{K}$ &&-0.34(07)(00)&&-0.37(06)(00)&&-0.08(15)(00)&-0.01(15)(00)&\\  
    \hline
    \hline
    \end{tabular}  
    \label{tab:final}
\end{table*}

Predictions for light meson masses and decay constants require non-perturbative renormalization and therefore introduce additional systematic uncertainties. In the previous CLQCD study~\cite{CLQCD:2023sdb}, the central values in the \(\overline{\mathrm{MS}}\) scheme were obtained via the RI/MOM scheme~\cite{Martinelli:1994ty}, and the difference between the RI/MOM and SMOM results was taken as a systematic uncertainty. Benefiting from the inclusion of data at three new lattice spacings, we are now able to reliably quantify the systematic uncertainty due to a potential \(\mathcal{O}(a\alpha_s)\) term. The renormalization constants obtained for both the CL@CL and CL@HI configurations are presented in the appendix.

Our final predictions--including quark masses, low-energy constants, pion and kaon decay constants, and other fit parameters--are summarized in Table~\ref{tab:final} for both the \(N_f=2+1\) unitary clover (CL@CL) setup and the \(N_f=2+1+1\) mixed action (CL@HI) setup. For comparison, the table also includes results from the previous CLQCD study~\cite{CLQCD:2023sdb}, which used a subset of the \(N_f=2+1\) CLQCD ensembles at three lattice spacings.

As shown in Table~\ref{tab:final}, the updated \(N_f=2+1\) CL@CL results are consistent with those from Ref.~\cite{CLQCD:2023sdb} when the same intermediate renormalization scheme (MOM or SMOM) is used. In certain cases, the uncertainties are slightly larger due to the inclusion of systematic uncertainty from possible \(a\alpha_s\) correction. The values of \(m_{l}\), \(m_s\), \(f_{\pi}\), and \(f_K\) obtained via the MOM scheme are consistent with those via the SMOM scheme within \(2\sigma\), and the scheme-independent ratio \(m_s/m_l\) is consistent with higher precision. 

For the \(N_f=2+1+1\) CL@HI results, the uncertainties are larger due to limited statistics. However, the discretization errors in \(f_{\pi}\) and \(f_K\) are significantly smaller, regardless of whether the MOM or SMOM scheme is used for renormalization. Consequently, the consistency between the results obtained with the two schemes is better than in the CL@CL case, although the impact of the larger statistical uncertainties warrants further investigation. Furthermore, the analysis shows that the systematic uncertainties arising from potential $a\alpha_s$ terms are considerably smaller for all the cases considered.
This observation provides additional evidence that our mixed action setup can also suppress discretization errors of hadronic matrix elements.

Given that the SMOM results have smaller renormalization uncertainties and already include the systematic uncertainty from possible \(a\alpha_s\) corrections, we adopt them as our final result.
The FLAG averages for both the 2+1 and 2+1+1 cases~\cite{FlavourLatticeAveragingGroupFLAG:2021npn,FlavourLatticeAveragingGroupFLAG:2024oxs,RBC:2012cbl,CLQCD:2023sdb,Bruno:2019vup,RBC:2014ntl,Laiho:2011np,BMW:2010ucx,BMW:2010skj,Bazavov:2010yq,McNeile:2010ji,RBC:2010qam,Davies:2009ih,MILC:2009ltw,MILC:2009mpl,Mason:2005bj,MILC:2004qnl,HPQCD:2004hdp,Bruno:2019xed,ExtendedTwistedMass:2021gbo,EuropeanTwistedMass:2014osg,FermilabLattice:2018est,Lytle:2018evc,Chakraborty:2014aca,Bazavov:2017lyh,FermilabLattice:2014tsy,ExtendedTwistedMass:2021qui,Miller:2020xhy,Carrasco:2014poa,Dimopoulos:2013qfa,Dowdall:2013rya,MILC:2013ffd,MILC:2011nja,Farchioni:2010tb,QCDSF-UKQCD:2016rau,Durr:2016ulb,Scholz:2016kcr,MILC:2010hzw,BMW:2010xmi,Aubin:2008ie,Follana:2007uv,Fahy:2015xka,Liang:2021pql,Alexandrou:2017bzk,Cichy:2013gja,Aoki:2017paw,Boyle:2015exm,BMW:2013fzj,Borsanyi:2012zv,Engel:2014eea,Brandt:2013dua,Burger:2012ti,ETM:2009ztk,Frezzotti:2008dr,ETM:2010cqp,Beane:2011zm,Horsley:2013ayv,Colangelo:2003hf,Gulpers:2015bba,JLQCD:2009ofg} are also listed in Table ~\ref{tab:final} for comparison.

\begin{figure}[hbt!]
\centering
\includegraphics[width=1.0\linewidth]{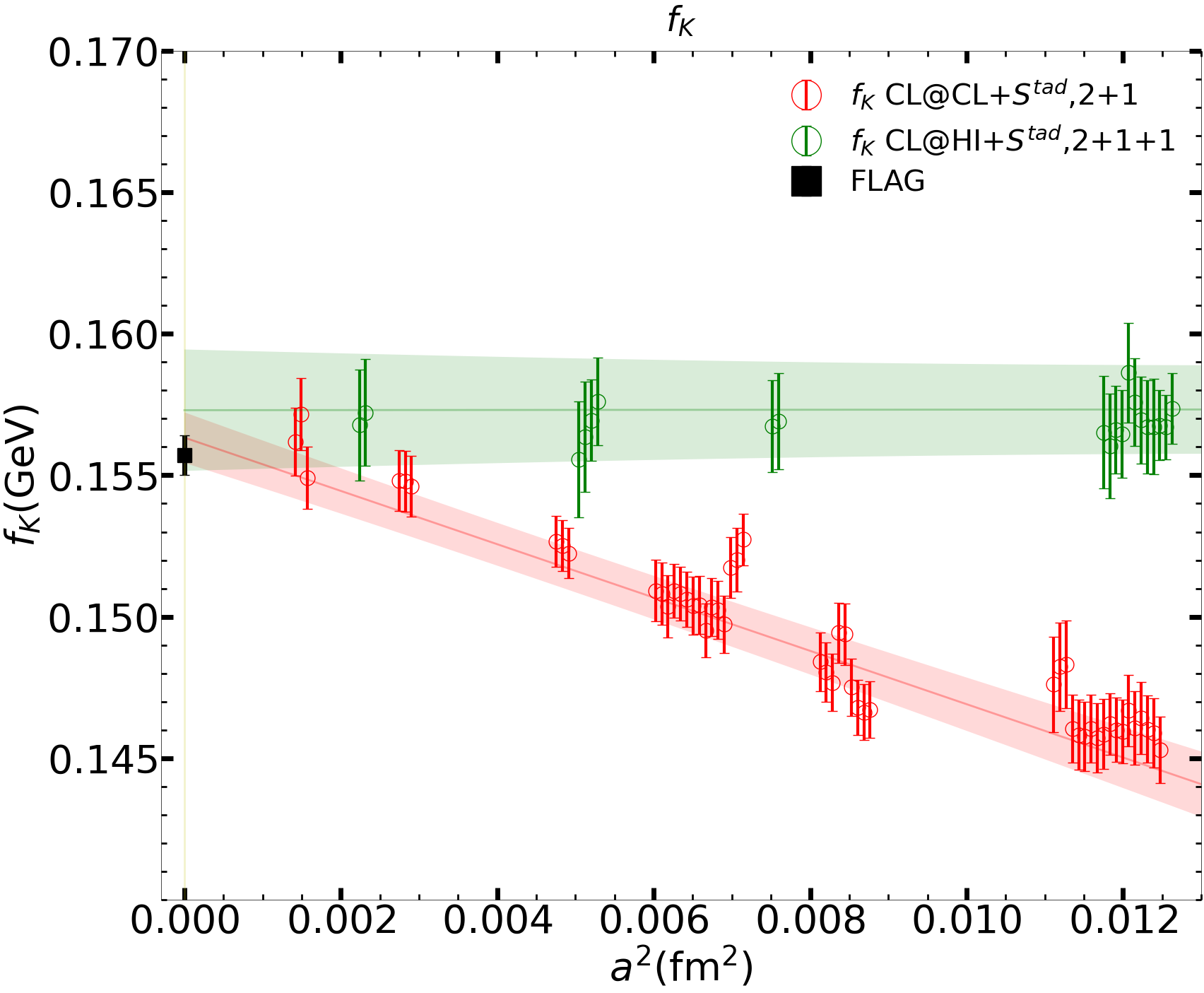}
\caption{Lattice spacing dependence of the $f_{K}$ with the impacts from the unphysical strange and light quark masses corrected using the parameters obtained from the joint fit.}\label{fig:fk}
\end{figure}

After correcting for unphysical strange- and light-quark masses using the parameters from the joint fit, we present the corrected kaon decay constant \( f_K \) through the RI/MOM renormalization, as a function of \( a^2 \) in Fig.~\ref{fig:fk}. The comparison between the \( N_f=2+1+1 \) (CL@HI) and \( N_f=2+1 \) (CL@CL) cases illustrates the suppression of discretization errors in hadronic matrix elements for the CL@HI setup.

\section{Summary}\label{sec:summary}

In summary, we studied the feature and discretization errors of hadron masses and meson decay constants using the tadpole-improved clover fermion with stout smeared gauge link, on the 2+1+1 flavor HISQ fermion ensembles. 
Compared to the similar calculation on the unitary 2+1 flavor clover fermion ensembles, the discretization errors of the hadronic observable likes $f_{\pi,K}$ and $m_{\phi,\Omega}$ are significantly suppressed. Thus even though the mixed action setup can introduce additional discretization effects, our calculation shows evidences that those effects can partially ``cancel" with the discretization error in the unitary clover setup, at least for light hadron observables, resulting in better control on the systematic uncertainty from the continuum extrapolation, using the combination of the stout smeared clover fermion on the HISQ sea with tadpole-improved Symanzik gauge action.

\begin{figure*}[hbt!]
\centering
\includegraphics[width=1.0\linewidth]{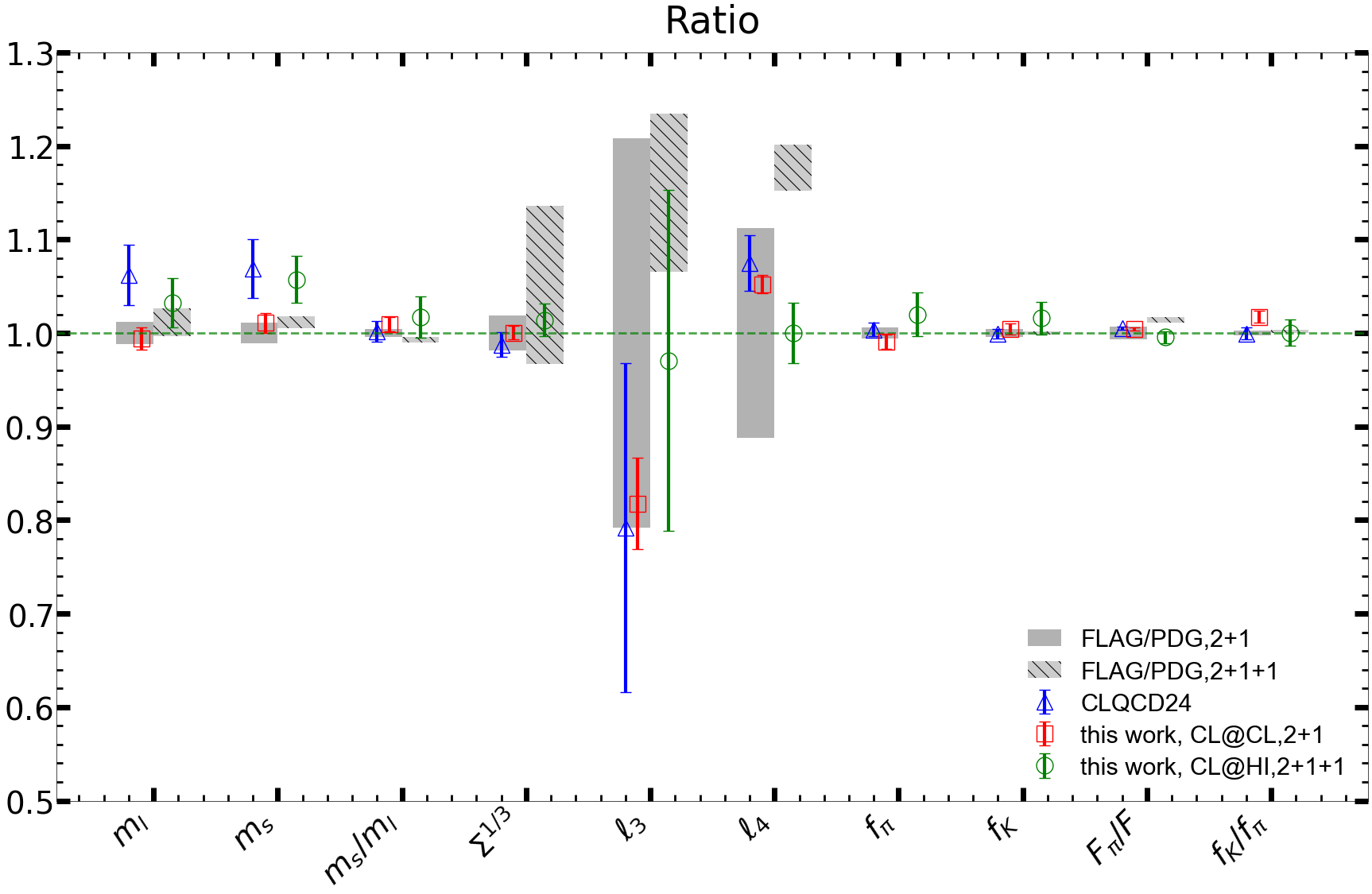}
\caption{Ratios of our predictions to the central values of the 2+1 flavor FLAG averages~\cite{FlavourLatticeAveragingGroupFLAG:2021npn,FlavourLatticeAveragingGroupFLAG:2024oxs}. All of our 2+1+1 flavor predictions (green) are consistent with the 2+1 ones (red) within $2\sigma$. Previous 2+1 flavor CLQCD results (blue) are also shown for comparison.}\label{fig:final}
\end{figure*}

Our investigation at $a\sim$ 0.1 fm further indicates that, within current statistical uncertainties, the inclusion of a dynamical charm quark sea does not alter low-energy hadronic physics, as illustrated in Fig.~\ref{fig:final} which normalizes all the quantities with the 2+1 flavor FLAG averages~\cite{FlavourLatticeAveragingGroupFLAG:2021npn,FlavourLatticeAveragingGroupFLAG:2024oxs}. Similar feature is also observed for the renormalization constants as detailed in the appendix. However, given the significantly smaller statistical errors achievable in charm-physics observable, such inclusion can modify the coefficients of the corresponding discretization errors. Furthermore, we find that hadron masses computed using the tadpole-improved Symanzik gauge action are consistent with those from the Iwasaki gauge action, agreeing well within uncertainties. Future studies at finer lattice spacings or with different valence fermion actions would provide more stringent tests of these findings.

\section*{Acknowledgement}
We thank the CLQCD collaborations for providing us their gauge configurations with dynamical fermions~\cite{CLQCD:2023sdb,CLQCD:2024yyn,Zhang:2025ieg}, which are generated on Advanced Computing East China Sub-center, HPC Cluster of ITP-CAS, IHEP-CAS and CSNS-CAS, ORISE Supercomputer, the Southern Nuclear Science Computing Center(SNSC) and the Siyuan-1 cluster supported by the Center for High Performance Computing at Shanghai Jiao Tong University. 
We thank Ying Chen, Hengtong Ding, Xu Feng, Chuan Liu, Liuming Liu, Wei Wang and the other CLQCD members for valuable comments and suggestions.
The calculations were performed using the PyQUDA package~\cite{Jiang:2024lto} with QUDA~\cite{Clark:2009wm,Babich:2011np,Clark:2016rdz} through HIP programming model~\cite{Bi:2020wpt}. The numerical calculation were carried out on the Advanced Computing East China Sub-center, ORISE Supercomputer and HPC Cluster of ITP-CAS. This work is supported in part by National Key R\&D Program of China No.2024YFE0109800, NSFC grants No. 12525504, 12435002, 12293060, 12293062, 12293061, 12293065 and 12447101, the Strategic Priority Research Program of Chinese Academy of Sciences, Grant No. YSBR-101.

\bibliography{ref}

\clearpage
\onecolumngrid

\section*{Appendix}


Table~\ref{tab:info_all} lists the tadpole factor \( v_0 \) (derived from smeared links for the clover term) and the tuned dimensionless valence clover masses \( \tilde{m}_{l,s,c}^{v} \). These masses are tuned to reproduce, respectively, the unitary pion mass, the physical \( \eta_s \) mass (\( 689.89(49)\,\text{MeV} \)), and the QED-subtracted \( D_s \) mass (\( 1966.7(1.5)\,\text{MeV} \)). For completeness, the table also includes the basic information for the HISQ ensembles, which have been shown in Table~\ref{tab:hisq_ensembles}.

\begin{table}[!h]
\centering
\caption{Summary table on the lattice spacing, $\tilde{L}^3\times \tilde{T}$, $m_{\pi}L$  and $m_{\pi,\eta_s,\eta_c}$ (in units of MeV) of the CLQCD ensembles with HISQ fermion~\cite{Zhang:2025ieg}. }
\begin{tabular}{l|lcc|ccc|cccc}
\hline
Ensemble & $a$ (fm) & $\tilde{L}^3 \times \tilde{T}$ &$m_{\pi}L$ & $m_{\pi}$ &$m_{\eta_s}$ & $m_{\eta_c}$ & $v_0$ & $\tilde{m}_l^{v}$ & $\tilde{m}_s^{v}$ & $\tilde{m}_c^{v}$  \\
\hline 
c24P31s &\multirow{6}[2]{*}{0.1084(4)}&$24^3\times 48$&4.13&313(2)&745(3)&2.973(12) &0.9636&-0.1946&-0.1557&\\
c24P31 &&$24^3\times 48$&4.07&309(2)&687(3)& 2.972(12)&0.9636&-0.1946&-0.1554&0.530 \\
c32P31 &&$32^3\times 48$&5.44&310(1)&686(3)&2.972(12)&0.9636&-0.1945&-0.1553&\\
c24P22 &&$24^3\times 48$&2.94&223(2)&685(3)&2.970(12) &0.9636&-0.1991&-0.1552&\\
c32P22 &&$32^3\times 48$&3.87&220(1)&684(3) &2.970(12)&0.9636&-0.1991&-0.1544&\\
c48P13 &&$48^3\times 48$&3.53&134(1)&683(3)& 2.970(12)&0.9636&-0.2016&-0.1540\\
\hline
e32P31 &0.0867(4)&$32^3\times 64$&4.41&313(2)&694(3)&3.015(13)&0.9678&-0.1645&-0.1352&0.350\\
\hline
g32P32 &\multirow{2}[2]{*}{0.0710(3)}&$32^3\times 64$&3.65&317(3)&692(3)&2.981(13)&0.9707&-0.1455&-0.1231&0.240\\
g48P31 &&$48^3\times 64$&5.38&311(2)&691(3)&2.983(13)&0.9707&-0.1457&-0.1231\\
\hline
h48P31 &0.0473(3)&$48^3\times 96$&3.60&313(3)&692(5)&2.947(19)&0.9754&-0.1173&-0.1035&0.106 \\
\hline
\hline
x24P31&0.1114(6)&$24^3\times 48$&4.35&321(2)&711(4)&--&0.9605& --&$-0.1755$&0.541\\
y24P31&0.1116(6)&$24^3\times 48$&4.22&311(2)&688(4)&--&0.9623&-0.2038&-0.1540&0.563\\
z24P31&0.1128(7)&$24^3\times 48$&4.35&317(2)&704(4)&--&0.9663&-- &$-0.1353$ &0.610\\
\hline
\end{tabular}
\label{tab:info_all}
\end{table}

Table~\ref{tab:CL@HI_Z} presents the renormalization constants \( Z_{X} \) for the currents \( X \equiv \bar{q}\Gamma_X q \), normalized by \( Z_V \) and matched to the \( \overline{\mathrm{MS}} \) scheme at 2 GeV via the RI/MOM and SMOM schemes, 
\begin{align}
 \frac{Z^{\rm MOM}_X(\mu^{\rm RI})}{Z_V} = \frac{\frac{1}{48}\text{Tr}[\Lambda^\mu_{V}(\mu^{\rm RI},0)\gamma_\mu]}{\frac{1}{12}\text{Tr}[\Lambda_{X}(\mu^{\rm RI},0)\Gamma_X]},\  \frac{Z^{\rm SMOM}_X(\mu^{\rm RI})}{Z_V} = \frac{\frac{1}{12q^2}\text{Tr}[q_{\mu}\Lambda^\mu_{V}(\mu^{\rm RI},1)q_{\nu}\gamma^\nu]}{\frac{1}{12}\text{Tr}[\Lambda_{X}(\mu^{\rm RI},1)\Gamma_X]}|_{q^2=(\mu^{\rm RI})^2},
\end{align}
for the clover valence fermion on the HISQ ensembles.
The regularization independent vertex functions with the momentum transfer $q\equiv p_1-p_2$ at the current is defined as,
\begin{align}
\Lambda_{\mathcal{O}}(\mu^{\rm RI},\omega) = S^{-1}(p_1) G_{\mathcal{O}}(p_1,p_2) S^{-1}(p_2) |_{p_1^2 = p_2^2 = (\mu^{\rm RI})^2,\ q^2=\omega (\mu^{\rm RI})^2}.
\end{align}

\begin{table}[!h]
\centering
\caption{Renormalization constants in $\overline{\mathrm{MS}}$ 2 GeV (if applicable) through the RI/MOM or SMOM scheme, for the clover fermion on the HISQ ensembles. The values of $Z_V$ are obtained from the pion matrix element which are the same in both schemes.}
\begin{tabular}{l|c|cc|cc|cc|cc}
\hline
Ensemble & $Z_V$ &\multicolumn{2}{c|}{ $Z_A/Z_V$} &\multicolumn{2}{c|}{$Z_S/Z_V$} & \multicolumn{2}{c|}{$Z_P/Z_V$} & \multicolumn{2}{c}{$Z_T/Z_V$} \\
& & MOM & SMOM & MOM & SMOM & MOM & SMOM & MOM & SMOM \\
\hline 
c24P31s &0.8476(2)&1.064(2) &1.077(6)&1.201(06)(38) &0.999(05)(04)&0.969(17)(28) &0.952(04)(04)&1.074(01)(05)&1.084(04)(12) \\
c24P31 &0.8476(2)&1.064(2) &1.079(5)&1.204(05)(35) &0.984(06)(04)&0.995(19)(28) &0.951(05)(04)&1.076(01)(05)&1.080(04)(12) \\
c32P31 &0.8479(1)&1.060(2) &1.072(5)&1.199(05)(35) &1.001(04)(04)&0.974(17)(27) &0.947(03)(04)& 1.072(01)(05)&1.081(03)(12) \\
c24P22 &0.8478(2)&1.061(2) &1.077(5)&1.193(04)(37) &0.997(05)(04)&0.965(14)(28) &0.953(03)(04)&1.074(01)(05)&1.085(04)(12)  \\
c32P22 &0.8481(1)&1.062(2) &1.080(5)&1.202(04)(36) &1.002(04)(04)&0.986(16)(27) &0.938(04)(04)&1.072(01)(05)&1.083(04)(12) \\
c48P13 &0.8484(1)&1.062(1) &1.071(3)&1.200(03)(35) &0.999(03)(04)&0.961(09)(26) &0.952(03)(04)&1.073(01)(05)&1.080(02)(12) \\
\hline
e32P31 &0.8720(1)&1.049(1) &1.060(2)&1.097(02)(23) &0.945(02)(04)&0.919(07)(19) &0.902(02)(04)&1.085(01)(04)&1.093(01)(10) \\
\hline
g32P32 &0.8872(2)&1.042(1) &1.050(1)&1.018(04)(21) &0.905(02)(05)&0.864(10)(17) &0.870(02)(05)&1.098(01)(04)&1.105(01)(09) \\
g48P31 &0.8876(1)&1.037(1) &1.042(1) &1.016(02)(23) &0.906(01)(05) &0.876(06)(17)&0.876(01)(05)&1.094(01)(04)&1.097(01)(09)\\
\hline
h48P31 &0.9108(1)&1.027(1) &1.032(1)&0.912(03)(11) &0.839(01)(06)&0.813(03)(10) &0.813(01)(06)&1.118(01)(03)&1.124(01)(07)\\
\hline
\hline
y24P31&0.8407(3)&1.063(2) &1.083(5)&1.212(11)(35) &1.005(06)(04)&0.966(34)(28) &0.958(04)(04)&1.072(02)(05)&1.082(04)(12)\\
\hline
\end{tabular}
\label{tab:CL@HI_Z}
\end{table}

Similar values for the unitary clover ensembles are collected in Table ~\ref{tab:CL@CL_Z}. Note that the SMOM scheme employed here is the same as that used for the bottom quark mass determination~\cite{Cai:2026xja}, but differs from the SMOM\(_{\gamma_{\mu}}\) scheme used in Ref.~\cite{CLQCD:2023sdb}. Consequently, the values presented in Table~\ref{tab:CL@CL_Z} deviate from those reported in Ref.~\cite{CLQCD:2023sdb}, although this discrepancy diminishes as the lattice spacing decreases.

In Table~\ref{tab:CL@HI_Z}, we also list the renormalization constants for the y24P31 ensemble at \(a = 0.112\) fm. This ensemble uses the same fermion and gauge actions as the others but omits the charm quark loop. Within uncertainties, all values are similar to those of the c24P31 ensemble at \(a = 0.108\) fm, with the exception of \(Z_V\), which exhibits a much smaller uncertainty. Extrapolating \(Z_V\) of the \(N_f = 2+1+1\) case to the lattice spacing of the \(N_f = 2+1+1\) y24P31 ensemble yields \(Z_V = 0.8419(3)\), which differs from the directly measured value \(Z_V = 0.8407(3)\) by only \(\sim 0.1\%\). This observation provides strong evidence for the insensitivity of the renormalization constants to the presence of heavy sea quarks.

\begin{table}[!h]
\centering
\caption{Similar to Table~\ref{tab:CL@HI_Z} but for the unitary clover fermion ensembles.}
\begin{tabular}{l|c|cc|cc|cc|cc}
\hline
Ensemble & $Z_V$ &\multicolumn{2}{c|}{ $Z_A/Z_V$} &\multicolumn{2}{c|}{$Z_S/Z_V$} & \multicolumn{2}{c|}{$Z_P/Z_V$} & \multicolumn{2}{c}{$Z_T/Z_V$} \\
& & MOM & SMOM & MOM & SMOM & MOM & SMOM & MOM & SMOM \\
\hline 
C24P34 &0.7968(3)&1.076(1)&1.094(4)&1.205(06)(41)&1.000(04)(04)&0.919(14)(33)&0.922(05)(04)&1.086(03)(06)&1.110(03)(12) \\
C24P29 &0.7981(2)&1.071(1)&1.089(3)&1.194(08)(39)&0.988(04)(04)&0.918(16)(30)&0.925(05)(04)&1.083(01)(06)&1.104(02)(12) \\
C32P29 &0.7981(1)&1.071(2)&1.085(3)&1.199(10)(43)&0.989(03)(04)&0.911(11)(33)&0.920(03)(04)&1.083(01)(06)&1.102(02)(12) \\
C32P23 &0.7996(1)&1.071(1)&1.081(3)&1.197(08)(41)&0.982(03)(04)&0.917(10)(33)&0.921(03)(04)&1.082(01)(06)&1.101(02)(12) \\
C48P23 &0.7995(1)&1.072(1)&1.094(3)&1.208(13)(43)&0.994(02)(04)&0.917(11)(34)&0.923(02)(04)&1.082(01)(06)&1.106(02)(12) \\
C48P14 &0.7996(1)&1.071(1)&1.088(3)&1.204(05)(43)&0.992(03)(04)&0.909(14)(33)&0.920(02)(04)&1.083(01)(06)&1.103(02)(12) \\
\hline
E28P35 &0.8177(1)&1.060(1)&1.071(1)&1.095(03)(20)&0.951(01)(04)&0.862(06)(17)&0.890(03)(04)&1.099(01)(04)&1.108(01)(11) \\
E32P29 &0.8188(1)&1.053(1)&1.068(2)&1.078(02)(21)&0.943(02)(04)&0.878(07)(17)&0.885(02)(04)&1.095(01)(04)&1.104(02)(10) \\
E32P22 &0.8196(1)&1.054(1)&1.060(3)&1.083(03)(22)&0.942(03)(04)&0.867(06)(18)&0.892(03)(04)&1.097(01)(05)&1.099(02)(11) \\
\hline
F32P30 &0.8355(1)&1.055(1)&1.059(2)&1.043(11)(26)&0.917(03)(05)&0.837(08)(20)&0.869(02)(04)&1.104(01)(05)&1.116(02)(10) \\
F48P30 &0.8351(1)&1.056(1)&1.061(2)&1.053(06)(22)&0.922(02)(04)&0.839(06)(18)&0.864(02)(04)&1.105(01)(04)&1.119(02)(10) \\
F32P21 &0.8358(1)&1.054(1)&1.058(2)&1.042(07)(25)&0.919(02)(05)&0.842(06)(20)&0.866(02)(04)&1.101(01)(05)&1.114(01)(10) \\
F48P21 &0.8357(1)&1.054(1)&1.066(3)&1.051(05)(20)&0.916(02)(05)&0.839(10)(18)&0.857(02)(04)&1.103(01)(04)&1.117(02)(10) \\
F64P14 &0.8359(1)&1.051(1)&1.065(2)&1.029(01)(17)&0.917(02)(05)&0.850(06)(14)&0.859(02)(04)&1.107(01)(03)&1.116(02)(10) \\
\hline
G36P29 &0.8464(1)&1.046(1)&1.051(1)&0.983(02)(20)&0.890(01)(05)&0.810(06)(16)&0.844(02)(05)&1.115(01)(04)&1.121(01)(09) \\
\hline
H48P32 &0.8681(1)&1.038(1)&1.043(1)&0.916(10)(12)&0.845(01)(05)&0.786(06)(09)&0.803(01)(05)&1.126(01)(02)&1.137(01)(08) \\
\hline
I64P31 &0.8892(1)&1.026(1)&1.029(1)&0.857(01)(08)&0.799(01)(06)&0.769(01)(07)&0.771(01)(06)&1.138(01)(04)&1.144(01)(06) \\
\hline
\end{tabular}
\label{tab:CL@CL_Z}
\end{table}

\begin{figure}[hbt!]
\centering
\includegraphics[width=0.45\linewidth]{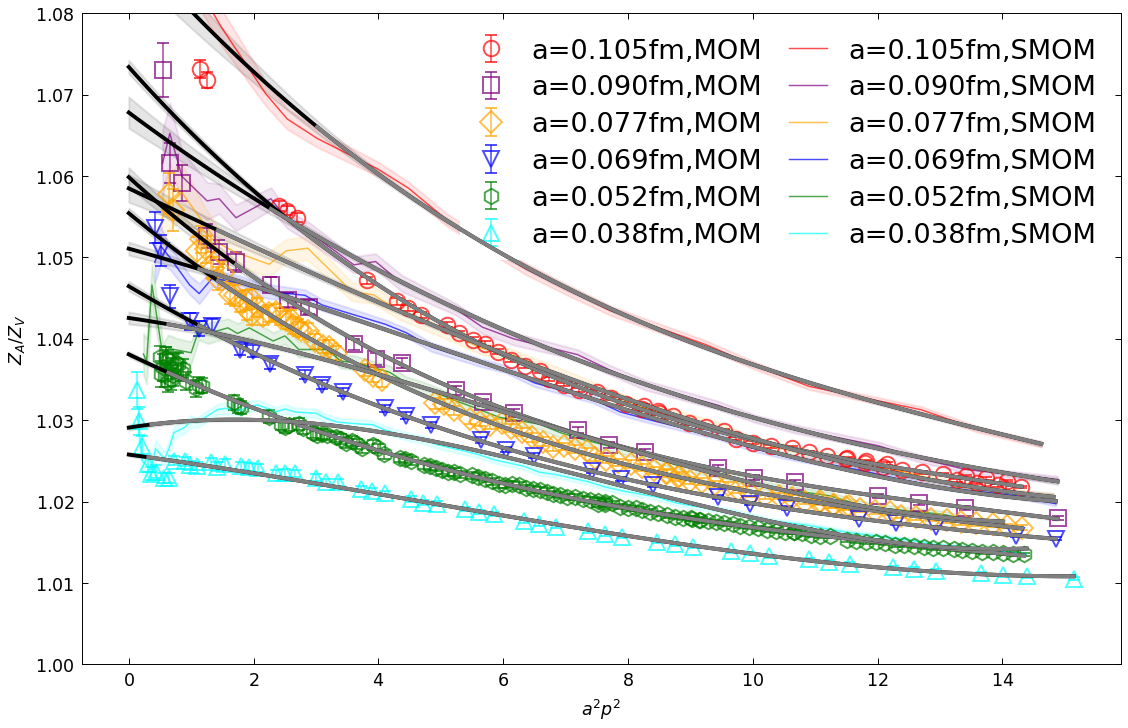}
\includegraphics[width=0.45\linewidth]{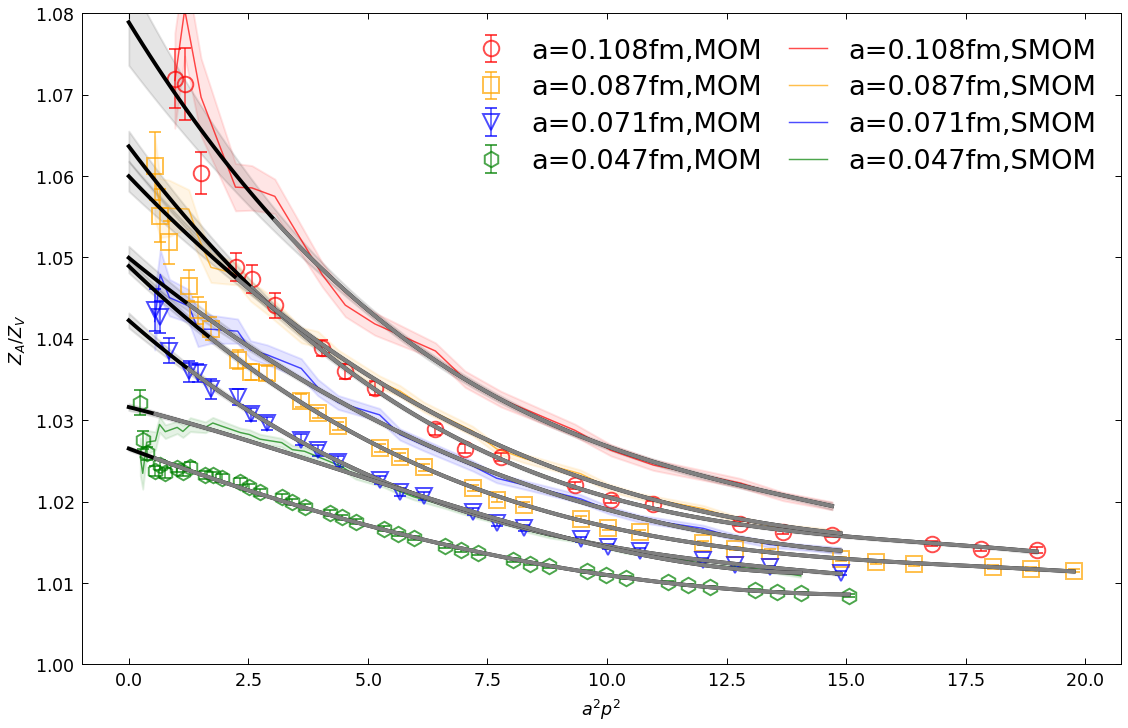}
\caption{\( a^2(\mu^{\rm RI})^2 \) dependence of \( Z_A/Z_V \) at different lattice spacing using the \( N_f=2+1 \) unitary clover setup (left panel) and the \( N_f=2+1+1 \) clover on HISQ setup (right panel).}\label{fig:ZA_ZV}
\end{figure}

Figure~\ref{fig:ZA_ZV} shows the ratio \( Z_A/Z_V \) calculated in the MOM (data points) and SMOM (bands) schemes at various lattice spacings \( a \), plotted as a function of \( a^2(\mu^{\rm RI})^2 \). This quantity characterizes the additive chiral symmetry breaking of the clover fermion action. Comparing the \( N_f=2+1 \) unitary clover case (CL@CL, left panel) with the \( N_f=2+1+1 \) clover-on-HISQ case (CL@HI, right panel), the extrapolated values of \( Z_A/Z_V \) to \( a^2(\mu^{\rm RI})^2 \rightarrow 0 \)--obtained from a polynomial fit to data with \( \mu^{\rm RI} \ge 3 \)~GeV--are similar at comparable lattice spacings.

\begin{figure}[hbt!]
\centering
\includegraphics[width=0.45\linewidth]{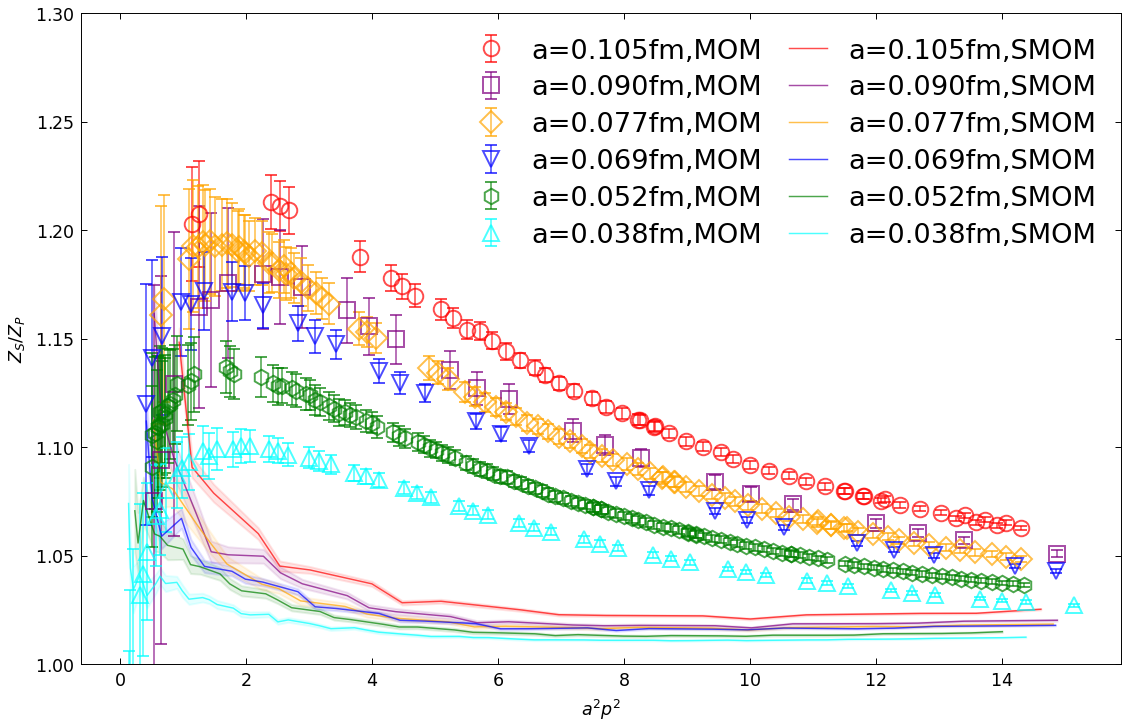}
\includegraphics[width=0.45\linewidth]{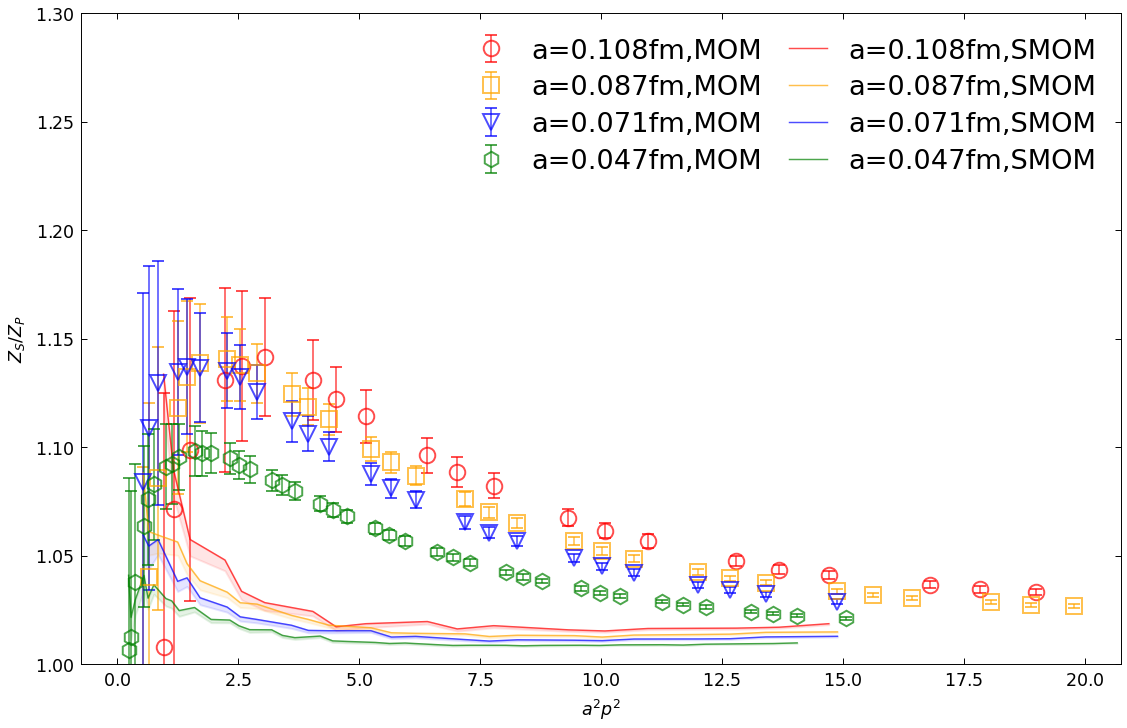}
\caption{Similar to Fig.~\ref{fig:ZA_ZV} but for $Z_S/Z_P$.}\label{fig:ZS_ZP}
\end{figure}

In contrast, the deviation of \( Z_S/Z_P \) from 1 is significantly larger in the CL@CL case than in the CL@HI case. As shown in Fig.~\ref{fig:ZS_ZP}, the difference is more pronounced at coarser lattice spacings. The suppression of \( Z_S/Z_P - 1 \) in the CL@HI setup likely originates from the smaller tadpole-improved bare coupling compared to the CL@CL case, a point that merits further investigation.

\end{document}